\documentstyle[fleqn,12pt,epsf]{article}
\newcommand{\tresj}[6]{\left( \begin{array}{ccc}
                              #1 & #2 & #3 \\
                              #4 & #5 & #6 
                             \end{array}
                      \right)}
\newcommand{\seisj}[6]{\left\{ \begin{array}{ccc}
                              #1 & #2 & #3 \\
                              #4 & #5 & #6 
                             \end{array}
                      \right\} }
\newcommand{\nuevej}[9]{\left\{ \begin{array}{ccc}
                              #1 & #2 & #3 \\
                              #4 & #5 & #6 \\
                              #7 & #8 & #9 
                             \end{array}
                      \right\} }

\newcommand{\Jb}{{\cal J}}
\newcommand{\Mb}{{\cal M}}

\newcommand{\nk}{{\bf k}}
\newcommand{\np}{{\bf p}}
\newcommand{\nq}{{\bf q}}

\newcommand{\hs}{\mbox{\boldmath{$\hat{s}$}}}
\newcommand{\ns}{\mbox{\boldmath{$s$}}}
\newcommand{\nl}{\mbox{\boldmath{$l$}}}
\newcommand{\nn}{\mbox{\boldmath{$n$}}}
\newcommand{\nt}{\mbox{\boldmath{$t$}}}

\newcommand{\nJ}{{\bf J}}

\newcommand{\nR}{{\bf R}}
\newcommand{\nP}{{\bf P}}
\newcommand{\hp}{{\bf \hat{p}}}

\newcommand{\nSigma}{\mbox{\boldmath$\Sigma$}}
\newcommand{\nDelta}{\mbox{\boldmath$\Delta$}}

\begin{document}
\begin{titlepage}
\thispagestyle{empty}
\mbox{} 
\vspace*{2.5\fill} 

{\Large\bf 
\begin{center}

Meson Exchange Currents in $(e,e'p)$ recoil polarization observables

\end{center}
} 

\vspace{1\fill} 

\begin{center}
{\large F. Kazemi Tabatabaei$^1$, J.E. Amaro$^1$ and J.A. Caballero$^2$}
\end{center}

{\small 
\begin{center}
$^1$ {\em Departamento de F\'{\i}sica Moderna, 
          Universidad de Granada,
          Granada 18071, Spain}  \\   
$^2$ {\em Departamento de F\'\i sica At\'omica, Molecular y Nuclear \\ 
          Universidad de Sevilla, Apdo. 1065, 
          Sevilla 41080, Spain 
}\\[2mm]
\end{center}
} 

\kern 1.5cm 
\hrule 
\kern 3mm 

{\small\noindent
{\bf Abstract} 
\vspace{3mm} 

A study of the effects of meson-exchange currents and isobar
configurations in $A(\vec{e},e'\vec{p})B$ reactions is presented. We
use a distorted wave impulse approximation (DWIA) 
model where final-state interactions are treated through a phenomenological
optical potential. The model includes relativistic corrections in the
kinematics and in the electromagnetic one- and two-body currents.  The full
set of polarized response functions is analyzed, as well as the transferred
polarization asymmetry. Results are presented for proton knock-out from closed-shell nuclei, for
moderate to high momentum transfer.
} 

\kern 3mm 
\hrule 

\end{titlepage}
\newpage
\setcounter{page}{1}


\section{Introduction}


 For the last decades coincidence (e,e$'$p) reactions on complex nuclei
have provided precise information on bound nucleon properties, which
have made it possible to test carefully the validity of present
nuclear models~\cite{Fru84,Bof93,Kel96,Bof96}.  Although the analysis
of these processes, making use of different distorted wave approaches
and coupled-channel models, has been extremely useful, there are still
uncertainties associated to the various ingredients that enter in the
description of the reaction mechanism: treatment of final-state
interactions (FSI), nuclear correlations, off-shell effects, Coulomb
distortion of the electrons, relativistic degrees of freedom,
meson-exchange currents (MEC), etc. All of these ingredients affect
the evaluation of the differential cross section and hence lead to
ambiguities in the extraction of the spectroscopic factors. The origin
of this uncertainty is directly connected with the complexity of the
dynamics of the reaction and the different approaches to handle it,
which produce different cross sections. It is clear that a reliable
determination of spectrospic factors requires an accurate description
of the reaction mechanism. Important efforts in this direction have
been done in recent works~\cite{Yim94,Gai00,Udi93,Udi01,Kel94}.

The measurement of the separate nuclear response functions and
asymmetries imposes additional restrictions over the theory. The
exclusive response functions, which include different components of
the hadronic tensor taken along the longitudinal (L) or transverse (T)
directions with respect to the momentum transfer $\nq$, may present
very different sensitivities to the different aspects of the
reaction. In this sense, it is interesting to point out that MEC are
shown to contribute mainly to the transverse
components~\cite{Ama93,Ama94,MEC}, while relativistic degrees of
freedom play a crucial role in the interference TL
response~\cite{Udi01,Udi99}.  Thus, a joint analysis of cross sections
and response functions, comparing the experimental data with the
theoretical predictions, can provide very relevant and complementary
information on the reaction mechanism.  Separate response functions
and the TL asymmetry have been measured for $^{16}$O(e,e$'$p) at
moderate~\cite{Chi91,Spa93} and high~\cite{Gao00} $q$-values.  The
asymmetry $A_{TL}$, obtained from the difference of cross sections
measured at opposite azimuthal angles (with respect to $\nq$) divided
by the sum, results particularly relevant because it does not depend
on the spectroscopic factors.  For high missing momentum $p\ge 300$
MeV/c, $A_{TL}$ presents an oscillatory structure that has been shown
to be consistent with predictions of `dynamical' relativistic
calculations~\cite{Udi99,Kel99b,Deb00,Cab01}.

The advent of longitudinally polarized beams~\cite{Man94} and recoil
polarization measurements~\cite{Woo98,Mal00} has importantly enlarged
the number of observables which can be accessible with this type of
experiments, a fact that is welcome to challenge the theory strongly. In
recent experiments carried out at MIT-Bates and Jefferson Lab, the
induced ($\nP$) and transferred ($\nP'$) polarization asymmetries were
measured for complex nuclei, $^{12}$C~\cite{Woo98} and
$^{16}$O~\cite{Mal00}, respectively.  In both cases
$(q,\omega)$-constant kinematics has been selected with $q\approx 760$
MeV/c, $\omega\approx 290$ MeV at MIT-Bates and $q\approx 1000$ MeV/c,
$\omega\approx 450$ MeV at TJlab. Since the transfer momentum values
are high enough, relativistic degrees of freedom should be
incorporated in a consistent description of these reactions. After the
pioneering work in~\cite{Pic87,Pic89}, a detailed study on the induced
normal polarization $P_n$ has been presented in~\cite{Joh99,Udi00}
within the framework of the relativistic distorted wave impulse
approximation (RDWIA).  A comparison with non-relativistic analyses
was also discussed.  The sensitivity of polarized observables to
channel coupling in final state interactions was analyzed
in~\cite{Kel96,Kel99}, while in~\cite{Kel99b} the study was focussed
on the effects of spinor distortion over the transfer polarization
ratio $P'_x/P'_z$.  In~\cite{Ito97} the whole eighteen recoil nucleon
polarized responses were computed from intermediate to high momentum
transfer in the Dirac eikonal formalism. A comparison between the
predictions of the Glauber and eikonal models for $P_n$ was presented
in~\cite{Deb02} with the aim of bridge the gap between the low- and
high-energy description of FSI. More recently a theoretical study of
kinematical and dynamical relativistic effects over polarized response
functions and polarization asymmetries has been performed in
\cite{Mar02a,Mar02b} within the relativistic plane-wave impulse
approximation (RPWIA). A general analysis of all the polarized
observables within the RDWIA is at present in progress~\cite{Cris03}.

Our main aim in this work is to explore in depth the role played by
the two-body currents in recoil nucleon polarization observables.
Some previous analyses on this subject have been done by the Pavia
group~\cite{Bof90,Bof91} and the Gent group~\cite{Ryc01,Ryc99}.  The
calculation of MEC in~\cite{Bof90,Bof91} makes use of an effective
one-body operator leading to results which, in the unpolarized case,
differ significantly from those obtained with other approaches that
describe properly the two-body currents~\cite{Slu94,Ama99}. Recently
the unpolarized model of \cite{Bof91} has been improved
in~\cite{Giu02}, but differences with other MEC calculations still
persist~\cite{Ama03b}.  In~\cite{Ryc99} the induced and transferred
polarization asymmetries $P_n$, $P'_l$ and $P'_t$ were evaluated for
different kinematical situations.  The model considered did not rely
on any empirical input with respect to the FSI, describing the bound
and scattering states as the solutions of the Schr\"odinger equation
with a mean field potential obtained from a Hartree-Fock calculation.
MEC were included based on the formalism developed in~\cite{Slu94}
which also differs from the MEC analysis performed
in~\cite{Ama99,Ama03b}. In addition, in~\cite{Ryc99} results for high
momentum transfer (up to $q=1$ GeV/c) were evaluated including
relativistic corrections into the one-body current operator obtained
through the Foldy-Wouthuysen method.

In this work we extend the DWIA+MEC model developed for unpolarized
reactions in Refs.~\cite{Ama99,Ama03b} in order to describe the spin
observables in $(\vec{e},e'\vec{p})$ processes from closed shell
nuclei. This model takes care of relativistic degrees of freedom by
making use of semi-relativistic (SR) operators for the one-body (OB)
current~\cite{Udi99,Maz02,Ama96a,Ama96b} as well as for the two-body
MEC~\cite{Ama03b,Ama98a,Ama02c,Ama03a}.  The SR currents are obtained
by a direct Pauli reduction of the corresponding relativistic
operators by expanding only in missing momentum over the nucleon mass
while treating the transferred energy and momentum exactly.
Relativistic kinematics for the ejected nucleon is assumed throughout
this work.  Finally FSI are incorporated through a phenomenological
optical potential which, for high momentum transfer, is taken as the
Schr\"odinger-equivalent form of a S-V Dirac optical potential.  The
goal of this work is to use the SR approach to evaluate the importance
of MEC effects upon the spin observables and their dependence on the
FSI for intermediate to high momentum transfer.  As a complete
relativistic distorted wave analysis of MEC in (e,e$'$p) processes is
still lacking ---the only study in this direction has been performed
taking into account only the contact current~\cite{Meu02}--- the use
of the SR model becomes, as a starting point, a convenient way of
implementing relativistic effects in existing non relativistic
descriptions of the reaction mechanism in order to explore the high
momentum region.

The paper is organized as follows: in Section 2 we outline the
DWIA formalism describing in detail the multipole
expansion done for the separate response functions. In Section 3 we
present our results for the polarized response functions and transferred
polarization asymmetries for selected kinematics near the quasielastic
peak. Finally our conclusions are drawn in Secion 4.


\section{DWIA model of $(\vec{e},e'\vec{p})$ }


\subsection{Cross section and response functions}

The general formalism for coincidence electron scattering on nuclei
involving polarization degrees of freedom has been presented in detail
in Refs.~\cite{Bof96,Pic87,Pic89,Ras89}.  In this section we simply
provide the basic description of our DWIA model focusing on the
development of the multipole expansion used to compute the response
functions. For this end we follow closely the multipole formalism
developed in~\cite{Ama98b} for polarized nuclei.

We consider the process depicted in Fig.~1, in which an incident
electron with four-momentum $K^\mu_e=(\epsilon_e,\nk_e)$ and helicity
$h$ interacts with a nucleus $A$, scatters through an angle $\theta_e$
to four-momentum $K'{}^\mu_e=(\epsilon',\nk'_e)$ and is detected in
coincidence with a nucleon with momentum $\np'$ and energy $E'$. The
four-momentum transferred to the nucleus is
$Q^\mu=K^\mu_e-K'{}^{\mu}_e=(\omega,\nq)$, verifying
$Q^2=\omega^2-q^2<0$. The polarization of the final nucleon is
measured along an arbitrary direction defined by the unitary vector
$\ns$. Assuming plane waves for the electrons and neglecting the
nuclear recoil, the cross section can be written in the extreme
relativistic limit (ERL) $m_e \ll \epsilon_e$, as~\cite{Ras89}
\begin{equation}
\frac{d\sigma}{d\epsilon'_e d\Omega'_e d\hp'}
= \Sigma + h \Delta \, ,
\label{cross-section}
\end{equation}
where a separation has been made into terms involving polarized and
unpolarized incident electrons.  Using the general properties of the
leptonic tensor it can be shown that both terms, $\Sigma$ and
$\Delta$, have the following decompositions:
\begin{eqnarray}
\Sigma 
&=& K \sigma_M \left( v_LR^L+v_T
    R^T+v_{TL}R^{TL}+v_{TT}R^{TT}\right),
\label{sigma}\\
\Delta 
&=&
K \sigma_M \left( v_{TL'}R^{TL'}+v_{T'}R^{T'}\right) \, ,
\label{delta}
\end{eqnarray}
$\sigma_M$ is the Mott cross section, the factor $K\equiv
m_Np'/(2\pi\hbar)^3$, with $m_N$ the nucleon mass, 
and the $v_\alpha$-coefficients are the usual
electron kinematical factors\footnote{In this work we consider the
kinematical factors similar to those expressions presented
in~\cite{Ras89}. Notice that these factors differ from the ones of
ref.~\cite{Pic89} in a global sign for $v_{TL}$, $v_{TL'}$ and
$v_{TT}$, and an additional $1/\sqrt{2}$ factor in the case of the
interference $TL$ coefficients}.

\begin{figure}[tb]
\begin{center}
\leavevmode
\def\epsfsize#1#2{1.0#1}
\epsfbox[190 480 430 700]{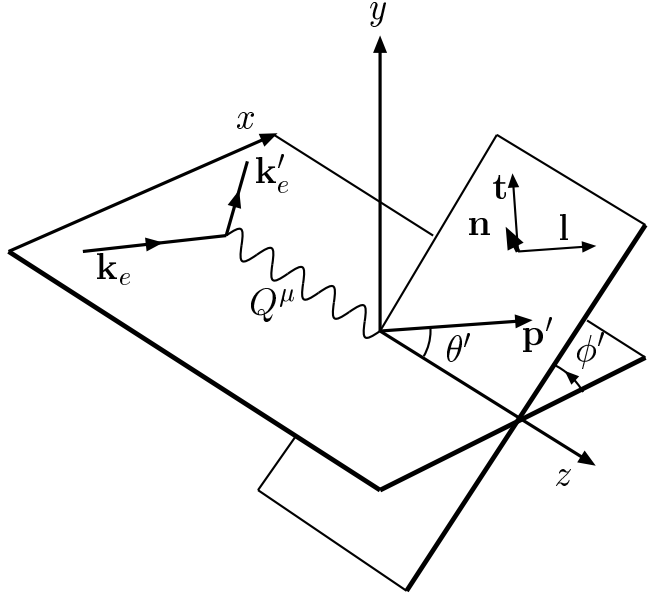}
\end{center}
\caption{Kinematics for the $(e,e'p)$ reaction. The $(x,y,z)$ coordinate
system is referred to the scattering plane with the
$z$-axis lying along the direction of the momentum transfer $\nq$. The
barycentric system $(l,t,n)$ is referred to the reaction plane:
$\nl$ lies along the direction of the ejected nucleon $\np'$,
the direction $\nn$ is defined by $\nq\times\np'$, and $\nt=\nn \times \nl$.
}
\end{figure}

The hadronic dynamics of the process is contained in the
exclusive response functions $R^K$, which are given as
\begin{eqnarray}
& R^L = W^{00}, 
& R^T = W^{xx}+W^{yy}, 
\label{rlrt}
\\ 
& R^{TL} = \sqrt{2}(W^{0x}+W^{x0}), 
& R^{TT} = W^{yy}-W^{xx}, 
\\ 
& R^{T'} =i(W^{xy}-W^{yx}), 
& R^{TL'} = i\sqrt{2}(W^{0y}+W^{y0}),
\label{rt'rtl'}
\end{eqnarray}
with $W^{\mu\nu}$ the hadronic tensor
\begin{equation}\label{hadronic-tensor}
W^{\mu\nu}
=\frac{1}{K}
\sum_{M_B}
\langle \np'\ns,B| \hat{J}^{\mu}(\nq)|A\rangle^*
\langle \np'\ns,B| \hat{J}^{\nu}(\nq)|A\rangle
\end{equation}
constructed from the matrix elements of the electromagnetic nuclear
current operator $\hat{J}^{\mu}(\nq)$ between the ground state of the
target nucleus $|A\rangle$ (assumed to have zero total angular
momentum), and the final hadronic states $|\np'\ns,B\rangle$. In what
follows we assume the residual nucleus to be left in a bound state,
hence its wave function can be written down in the form
$|B\rangle=|J_B M_B\rangle$ with $J_B$ the total angular momentum. The
state $|\np'\ns\rangle$ represents the asymptotic distorted wave
function of the ejected nucleon polarized along an arbitrary
$\ns$-direction, determined by the angles ($\theta_s,\phi_s)$ referred
to the $xyz$ coordinate system of Fig.~1. It is given by
\begin{equation}
|\np'\ns\rangle = 
\sum_{\nu=-1/2}^{1/2}{\cal D}_{\nu\frac12}^{(1/2)}(\theta_s,\phi_s,0)\
|\np' \nu\rangle \, ,
\label{rotation}
\end{equation}
where $|\np' \nu\rangle$ is referred to the system with the
quantization axis along $\nq$ and the arguments of the rotation
matrices are the Euler angles that specify the $\ns$-direction.

Isolating the explicit dependences on the azimuthal angle of the
ejected nucleon $\phi'=\phi$, the hadronic responses can be expressed
in the form
\begin{eqnarray}
R^L   &=& W^L  \label{RL}\\
R^T   &=& W^T  \\
R^{TL}&=& \cos\phi\ W^{TL} + \sin\phi\ \widetilde{W}^{TL}\\
R^{TT}&=& \cos2\phi\ W^{TT} + \sin2\phi\ \widetilde{W}^{TT}\\
R^{T'}&=& \widetilde{W}^{T'} \label{RT'}\\
R^{TL'}&=& \cos\phi\ \widetilde{W}^{TL'}+\sin\phi\ W^{TL'} \label{RTL'} \, ,
\end{eqnarray}
where the functions $W^K$ and $\widetilde{W}^K$ are totally specified
by four kinematical variables, for instance $\{E,\omega,q,\theta'\}$,
and the polarization direction $\{\theta_s,\Delta\phi=\phi-\phi_s\}$.
The responses with and without tilde refer to their dependence on the
spin vector $\ns$.  As shown below, $\widetilde{W}^K$ are purely
spin-vector, while $W^K$ present also a spin-scalar dependence, so
only the latter survive when the polarization of the ejected nucleon
is not measured.

In the case of $(\vec{e},e'\vec{N})$ processes, the hadronic response
functions are usually given by referring the recoil nucleon
polarization vector $\ns$ to the baryocentric system defined by the
axes (see Fig.~1): $\nl$ (along the $\np'$ direction), $\nn$ (normal
direction to the plane defined by $\nq$ and $\np'$, i.e., along
$\nq\times\np'$) and $\nt$ (determined by
$\nn\times\nl$)\footnote{Notice that this notation does not coincide
with Refs.~\cite{Mar02a,Mar02b} where the $\nt$ direction is denoted
as $\ns$ (sideways)}. It can be shown (see Refs.~\cite{Pic87,Pic89}
for details) that a total of eighteen response functions enter in the
analysis of $(\vec{e},e'\vec{N})$ reactions. These are given by the
decomposition
\begin{eqnarray}
W^K 
&=&
 \frac12 W_{unpol}^K +  W_n^K s_n, 
\kern 1cm  K=L,T,TL,TT,TL',
\label{Wgeneral}
\\
\widetilde{W}^{K}  
&=&
 W_l^{K}s_l + W_t^{K}s_t, \kern 1cm  K=TL,TT,T',TL' \, ,
\label{Wtgeneral}
\end{eqnarray}
where, as mentioned above, only the $W_{unpol}^K$ responses survive
within the unpolarized case.  Moreover, $W_{unpol}^{TL'}$ (referred as
fifth response) enters only when the polarization of the incident
electron is measured.

Owing to the above decomposition, the response
functions~(\ref{RL}--\ref{RTL'}) can be expressed in the form
$R^K=R^K_{unpol}/2 +\nR^K\cdot\ns$, and similarly, the cross
section~(\ref{cross-section}--\ref{delta}) can be written as a sum of
unpolarized and spin-vector dependent terms
\begin{eqnarray}
\frac{d\sigma}{d\epsilon'_e d\Omega'_e d\hp'}
&=&
 \frac12\Sigma_{unpol} + \nSigma\cdot\ns
+h\left( \frac12\Delta_{unpol} + \nDelta\cdot\ns\right)
\\
&=&
\frac12\Sigma_{unpol}\left[1+\nP\cdot\ns+h( A + \nP'\cdot\ns)\right] \, ,
\end{eqnarray}
where the usual polarization asymmetries have been introduced~\cite{Kel96}:
\begin{eqnarray}
\nP &=& \nSigma/\left(\frac12\Sigma_{unpol}\right)
\kern 1cm \mbox{Induced polarization},
\\
\nP' &=& \nDelta/\left(\frac12\Sigma_{unpol}\right)
\kern 1cm \mbox{Transferred polarization,}
\\
A &=& \Delta_{unpol}/\Sigma_{unpol}
\kern 1cm \mbox{Electron analyzing power.}
\end{eqnarray}

\subsection{Multipole analysis of  response functions}

In this section we present the multipole expansion of the response
functions to be used in our DWIA model. The final expressions, where
the sums over third components of angular momenta have been performed
analytically, result convenient in the present work since the
computational time can be considerably reduced, specially the
calculation concerning the MEC. Note that the number of multipoles
needed to get convergence increases with $q,\omega$ and up to $\sim
36$ multipoles are needed for $q=1$ GeV/c. The expansion is performed
following the formalism developed in~\cite{Ama98b} for exclusive
reactions from polarized nuclei. A basic difference
between the present work and that of ref.~\cite{Ama98b} lies on the
sums performed over the third components which are different when
initial and/or final state polarizations are considered. Here we
simply present the final expressions, referring to ref.~\cite{Ama98b}
for details on the expansion method and to the Appendix for an outline
on the procedure used to perform the sum over third components in the
present case.

In order to compute the hadronic tensor in our DWIA model we first perform a
multipole expansion of the ejected nucleon wave function in partial
waves. The final hadronic states may then be written 
\begin{equation}
|\np' \nu, B\rangle = 
\sum_{lMjm_pJ_fM_f}
i^l Y^*_{lM}(\hp')
\langle {\textstyle \frac12}\nu l M| jm_p\rangle
\langle j m_p J_BM_B|J_f M_f\rangle
|(lj)J_B,J_fM_f\rangle \, ,
\label{final-states}
\end{equation}
where the partial waves $(lj)$ are coupled to the angular momentum 
$J_B$ of the residual nucleus to give a total angular momentum $J_f$ 
in the final states $|f\rangle=|(lj)J_B,J_fM_f\rangle$. 

The electromagnetic charge and transverse current
operators are expanded as sums involving Coulomb (C), electric (E) and
magnetic (M) tensor operators,
\begin{eqnarray}
\hat{\rho}(q) &=& \sqrt{4\pi}\sum_{J=0}^{\infty} i^J [J] \hat{M}_{J0}(q)
\label{m-rho}
\\
\hat{J}_{m} &=& -\sqrt{2\pi}\sum_{J=1}^{\infty} i^J [J] 
            \left[\hat{T}^{el}_{Jm}(q)+m\hat{T}^{mag}_{Jm}(q)
            \right] \, ,
\kern 1cm  m=\pm 1 \, ,
\label{m-J}
\end{eqnarray}
where, as usual, we assume the transfer momentum $\nq$ along the
$z$-direction and $\hat{J}_{m}$ are the spherical components of the
current operator $\hat{\nJ}$. We use the bracket symbol
$[J]=\sqrt{2J+1}$ for angular
momenta. Inserting~(\ref{rotation},\ref{final-states},\ref{m-rho},\ref{m-J})
into the hadronic tensor (\ref{hadronic-tensor}), the following
expansion for the responses $W^K$ and $\widetilde{W}^K$ is obtained
\begin{equation}
W^K =
\frac12 W^K_{unpol}
+2\pi  P_1^{1}(\cos\theta_s) \sin(\Delta\phi) 
\widetilde{W}^{K}_{11},
\label{Wexpansion}
\end{equation}
for $K=L,T,TL,TT,TL'$, and 
\begin{equation}
\widetilde{W}^K
=
2\pi \alpha_K \left[  P_1^{0}(\cos\theta_s)W^{K}_{10}
           +  P_1^{1}(\cos\theta_s) \cos(\Delta\phi)W^{K}_{11} 
\right]
\label{Wtexpansion}
\end{equation}
for $K=TL,TT,T',TL'$, with the coefficient $\alpha_K=-1$ for $TL,TT$
and $\alpha_K=1$ for $T',TL'$, and 
$P_{\Jb}^{\Mb}(\cos\theta_s)$ the Legendre functions.

The five response functions $W^K_{unpol}$ are the only ones that survive
when summing over final spins $\pm \ns$, in which case
the 1/2 factor cancels and the unpolarized cross
section is recovered. The spin dependence is determined from the thirteen
reduced response functions $W^K_{1\Mb}$ ($\Mb=0,1$) and
$\widetilde{W}^K_{11}$ introduced above.
Explicit expressions for
these reduced responses can be written in terms
of the reduced matrix elements of the current multipole operators
\begin{eqnarray}
C_\sigma &=& \langle (lj)J_B,J\|\hat{M}_J\|0\rangle
\label{Csigma}
\\
E_\sigma &=& \langle (lj)J_B,J\|\hat{T}^{el}_J\|0\rangle
\label{Esigma}
\\
M_\sigma &=& \langle (lj)J_B,J\| i\hat{T}^{mag}_J\|0\rangle \, ,
\label{Msigma}
\end{eqnarray}
where we have defined a multiple index $\sigma = (l,j,J)$
corresponding to the quantum numbers of the final states. Note that
the initial state $|A\rangle=|0\rangle$ has total angular momentum
equal to zero, so $J_f=J$.  The response functions involve quadratic
products of these multipole matrix elements which can be decomposed
into their real ($R^K_{\sigma'\sigma}$) and imaginary
($I^K_{\sigma'\sigma}$) parts:
\begin{eqnarray}
C^*_{\sigma'}C_{\sigma} 
&=&
 R^L_{\sigma'\sigma}+iI^L_{\sigma'\sigma}
\label{ccL}
\\
E^*_{\sigma'}E_{\sigma} 
+M^*_{\sigma'}M_{\sigma} 
&=&
 R^{T1}_{\sigma'\sigma}+iI^{T1}_{\sigma'\sigma}
\\
E^*_{\sigma'}M_{\sigma} 
-M^*_{\sigma'}E_{\sigma} 
&=&
 R^{T2}_{\sigma'\sigma}+iI^{T2}_{\sigma'\sigma}
\\
C^*_{\sigma'}E_{\sigma} 
&=&
 R^{TL1}_{\sigma'\sigma}+iI^{TL1}_{\sigma'\sigma}
\\
C^*_{\sigma'}M_{\sigma} 
&=&
 R^{TL2}_{\sigma'\sigma}+iI^{TL2}_{\sigma'\sigma}
\\
E^*_{\sigma'}E_{\sigma} 
-M^*_{\sigma'}M_{\sigma} 
&=&
 R^{TT1}_{\sigma'\sigma}+iI^{TT1}_{\sigma'\sigma}
\\
E^*_{\sigma'}M_{\sigma} 
+M^*_{\sigma'}E_{\sigma} 
&=& R^{TT2}_{\sigma'\sigma}+iI^{TT2}_{\sigma'\sigma} \, .
\label{emTT2}
\end{eqnarray}
Expressions for the unpolarized response 
functions $W^K_{unpol}$ in terms of (\ref{ccL}--\ref{emTT2}) 
are given in ref.~\cite{Maz02}, while the recoil nucleon polarized responses
can be written as
\begin{eqnarray}
\widetilde{W}^L_{11}
&=&
\frac{1}{K} 
\sum_{\Jb'L}\tilde{h}^1_{1\Jb'L0}(\theta')
\sum_{\sigma'\sigma}P^+_{l+l'+\Jb'}\Phi_{\sigma'\sigma}
\tresj{J'}{J}{L}{0}{0}{0}
\xi^+_{J'-l',J-l}I^L_{\sigma'\sigma}
\label{WL11}
\\
\widetilde{W}^T_{11}
&=&
-\frac{1}{K} 
\sum_{\Jb'L} P^+_{\Jb'+L} \tilde{h}^1_{1\Jb'L0}(\theta')
\sum_{\sigma'\sigma}P^+_{l+l'+\Jb'}\Phi_{\sigma'\sigma}
\tresj{J'}{J}{L}{1}{-1}{0}
\nonumber\\
&&\mbox{}\times
\left(\xi^+_{J'-l',J-l}I^{T1}_{\sigma'\sigma}
     +\xi^-_{J'-l',J-l}I^{T2}_{\sigma'\sigma}
\right)
\\
\widetilde{W}^{TL}_{11}
&=&
-\frac{1}{K} 
2\sqrt{2}\sum_{\Jb'L}(-1)^{\Jb'+L}\tilde{h}^1_{1\Jb'L1}(\theta')
\sum_{\sigma'\sigma}P^+_{l+l'+\Jb'}\Phi_{\sigma'\sigma}
\tresj{J'}{J}{L}{0}{1}{-1}
\nonumber\\
&&\mbox{}\times
\left(\xi^+_{J'-l',J-l}I^{TL1}_{\sigma'\sigma}
     -\xi^-_{J'-l',J-l}I^{TL2}_{\sigma'\sigma}
\right)
\\
\widetilde{W}^{TT}_{11}
&=&
-\frac{1}{K} 
\sum_{\Jb'L}(-1)^{\Jb'+L}\tilde{h}^1_{1\Jb'L2}(\theta')
\sum_{\sigma'\sigma}P^+_{l+l'+\Jb'}\Phi_{\sigma'\sigma}
\tresj{J'}{J}{L}{1}{1}{-2}
\nonumber\\
&&\mbox{}\times
\left(\xi^+_{J'-l',J-l}I^{TT1}_{\sigma'\sigma}
     -\xi^-_{J'-l',J-l}I^{TT2}_{\sigma'\sigma}
\right)
\\
\widetilde{W}^{TL'}_{11}
&=&
-\frac{1}{K} 
2\sqrt{2}\sum_{\Jb'L}(-1)^{\Jb'+L}\tilde{h}^1_{1\Jb'L1}(\theta')
\sum_{\sigma'\sigma}P^+_{l+l'+\Jb'}\Phi_{\sigma'\sigma}
\tresj{J'}{J}{L}{0}{1}{-1}
\nonumber\\
&&\mbox{}\times
\left(\xi^+_{J'-l',J-l}R^{TL1}_{\sigma'\sigma}
     -\xi^-_{J'-l',J-l}R^{TL2}_{\sigma'\sigma}
\right)
\\
{W}^{TL}_{1\Mb}
&=&
-\frac{1}{K} 
2\sqrt{2}\sum_{\Jb'L}(-1)^{\Jb'+L}h^{\Mb}_{1\Jb'L1}(\theta')
\sum_{\sigma'\sigma}P^+_{l+l'+\Jb'}\Phi_{\sigma'\sigma}
\tresj{J'}{J}{L}{0}{1}{-1}
\nonumber\\
&&\mbox{}\times
\left(\xi^+_{J'-l',J-l}I^{TL1}_{\sigma'\sigma}
     -\xi^-_{J'-l',J-l}I^{TL2}_{\sigma'\sigma}
\right)
\\
{W}^{TT}_{1\Mb}
&=&
-\frac{1}{K} 
\sum_{\Jb'L}(-1)^{\Jb'+L}{h}^{\Mb}_{1\Jb'L2}(\theta')
\sum_{\sigma'\sigma}P^+_{l+l'+\Jb'}\Phi_{\sigma'\sigma}
\tresj{J'}{J}{L}{1}{1}{-2}
\nonumber\\
&&\mbox{}\times
\left(\xi^+_{J'-l',J-l}I^{TT1}_{\sigma'\sigma}
     -\xi^-_{J'-l',J-l}I^{TT2}_{\sigma'\sigma}
\right)
\\
{W}^{T'}_{1\Mb}
&=&
-\frac{1}{K} 
\sum_{\Jb'L} P^-_{\Jb'+L} {h}^{\Mb}_{1\Jb'L0}(\theta')
\sum_{\sigma'\sigma}P^+_{l+l'+\Jb'}\Phi_{\sigma'\sigma}
\tresj{J'}{J}{L}{1}{-1}{0}
\nonumber\\
&&\mbox{}\times
\left(\xi^+_{J'-l',J-l}R^{T1}_{\sigma'\sigma}
     +\xi^-_{J'-l',J-l}R^{T2}_{\sigma'\sigma}
\right)
\\
{W}^{TL'}_{1\Mb}
&=&
-\frac{1}{K} 
2\sqrt{2}\sum_{\Jb'L}(-1)^{\Jb'+L}{h}^{\Mb}_{1\Jb'L1}(\theta')
\sum_{\sigma'\sigma}P^+_{l+l'+\Jb'}\Phi_{\sigma'\sigma}
\tresj{J'}{J}{L}{0}{1}{-1}
\nonumber\\
&&\mbox{}\times
\left(\xi^+_{J'-l',J-l}R^{TL1}_{\sigma'\sigma}
     -\xi^-_{J'-l',J-l}R^{TL2}_{\sigma'\sigma}
\right) \, ,
\label{WTL'1M}
\end{eqnarray}
where we use the parity functions
\begin{equation}
P^{\pm}_{n}=(1\pm (-1)^n)/2, 
\kern 0.3cm
\xi^+_{J'J}\equiv (-1)^{(J'-J)/2}P^+_{J'+J},
\kern 0.3cm
\xi^-_{J'J}\equiv (-1)^{(J'-J+1)/2}P^-_{J'+J}
\end{equation}
and the angular dependence of the above responses is determined by the
functions $h^{\Mb}_{\Jb\Jb'LM}(\theta')$ and
$\tilde{h}^{\Mb}_{\Jb\Jb'LM}(\theta')$, defined through the coupling
of two spherical harmonics (see~(\ref{Ycoupling})).
\begin{eqnarray}
{\rm Re}\left[ Y_{\Jb}(\ns)Y_{\Jb'}(\hp')\right]_{LM}
&=&
\cos M\phi'
\sum_{\Mb=0}^{\Jb}
h^{\Mb}_{\Jb\Jb'LM}(\theta')
P^{\Mb}_{\Jb}(\cos\theta_s)\cos(\Mb\Delta\phi)
\nonumber\\
&&
\mbox{}+
\sin M\phi'
\sum_{\Mb=0}^{\Jb}
\tilde{h}^{\Mb}_{\Jb\Jb'LM}(\theta')
P^{\Mb}_{\Jb}(\cos\theta_s)\sin(\Mb\Delta\phi) \, .
\end{eqnarray}
Finally, the coefficients $\Phi_{\sigma'\sigma}$ are derived in the
Appendix and are given by eq.~(\ref{phi}) selecting $\Jb=1$. Although
the above expressions correspond formally to those denoted as
$W^{K(-)}_{\Jb\Mb}$ and $\widetilde{W}^{K(-)}_{\Jb\Mb}$
in~\cite{Ama98b} for polarized nuclei and $\Jb=1$, it is important to
point out that the coefficients $\Phi_{\sigma'\sigma}$ contain the
whole information on the polarization distribution of the
particles. Hence the significance of $\Phi_{\sigma'\sigma}$ is clearly
different when polarization degrees of freedom are considered for the
ejected nucleon (present work) or the target nucleus~\cite{Ama98b}.

The nuclear structure information in~(\ref{WL11}-\ref{WTL'1M}) 
is contained in the quadratic forms (\ref{ccL}--\ref{emTT2}) 
of the $C$, $E$, $M$ multipoles%
\footnote{Note that there is a typo in eqs.~(40--51) of
ref.~\cite{Ama98b}: the order of $J$ and $J'$ in the three-j should be
reversed. This error has been corrected in eqs.~(37--45) in the
present paper.}.  Thus the present expansion can be applied to any
nuclear model of the reaction as far as it provides multipole matrix
elements (\ref{Csigma},\ref{Esigma},\ref{Msigma}) for high enough
angular momenta $\sigma=(l,j,J)$. Note that only the responses
involving the real parts $R^K_{\sigma'\sigma}$ survive when FSI are
neglected since in this case all the $C$, $E$, $M$ multipoles are
strictly real functions. Therefore those responses which depend on the
imaginary parts are expected to be particularly sensitive to the
description of FSI.

Writing down explicitly the Legendre polynomials involved in 
the multipole expansion (\ref{Wexpansion},\ref{Wtexpansion}), and
comparing with the general expression (\ref{Wgeneral},\ref{Wtgeneral}),
we get the following relation between both sets of response functions:
\begin{eqnarray}
W^K_n 
&=& 
-2\pi \widetilde{W}^K_{11}, \kern 5cm K=L,T,TL,TT,TL'
\label{WKn}
\\
W^{K}_l 
&=& 
2\pi\alpha_K\left(W^{K}_{11}\sin\theta'+W^{K}_{10}\cos\theta'\right),
\kern 1cm K=TL,TT,TL',T'
\label{WKl}
\\
W^{K}_t 
&=& 
2\pi\alpha_K\left(W^{K}_{11}\cos\theta'-W^{K}_{10}\sin\theta'\right),
\kern 1cm K=TL,TT,TL',T'
\label{WKt}
\end{eqnarray}
with $\alpha_K$ as introduced in~(\ref{Wtexpansion}).

\subsection{Electromagnetic operators and PWIA}

In this work we evaluate the exclusive polarized responses using a
semi-relativistic (SR) model for describing the electromagnetic
one-body (OB) and two-body MEC current operators. The OB current has
been obtained by a direct Pauli reduction of the fully relativistic
operator in powers only of the initial nucleon momentum over the
nucleon mass $\np/m_N$. The dependence on the transfer and final
momenta, which can be large~\cite{Ama96a,Ama96b,Ama98a,Ama02c}, is
treated exactly.  The SR-OB current in momentum space can be written
as
\begin{eqnarray}
J^0(\np',\np) 
&=& 
\rho_c+ i \rho_{so}(\cos\phi\ \sigma_x-\sin\phi\ \sigma_y)\chi
\label{rho}
\\
J^x(\np',\np)
&=&
iJ_m\sigma_y + J_c\ \chi\cos\phi
\label{Jx}
\\
J^y(\np',\np)
&=&
-iJ_m\sigma_x + J_c\ \chi\sin\phi \, ,
\label{Jy}
\end{eqnarray}
where $\chi=(p/m_N)\sin\theta$, and $(\theta,\phi)$ are the angles
determining the direction of the initial momentum $\np$ in the
$(x,y,z)$ coordinate system.  The charge ($\rho_c$), spin-orbit
($\rho_{so}$), magnetization ($J_m$) and convection ($J_c$) terms
shown above include relativistic corrections and are given by the
following expressions
\begin{eqnarray}
\rho_c = \frac{\kappa}{\sqrt{\tau}}G_E, &&
\rho_{so} = \kappa\frac{2G_M-G_E}{2\sqrt{1+\tau}}
\label{rho-factors}
\\
J_m = \sqrt{\tau}G_M ,
&&
J_c = \frac{\sqrt{\tau}}{\kappa}G_E \, ,
\label{J-factors}
\end{eqnarray}
where $\kappa=q/2m_N$, $\tau=|Q^2|/4m_N^2$, and $G_E$, $G_M$ are the
electric and magnetic nucleon form factors for which we take the
Galster parameterization~\cite{Gal71}.

The two-body MEC operators of pionic (P), seagull or contact (S) and
$\Delta$-isobar kinds, displayed in the Feynman diagrams of Fig.~2,
have been also obtained by making use of a SR approach leading to
simple prescriptions that include relativistic corrections through a
multiplicative factor (see Refs.~\cite{Ama03b,Ama02c,Ama03a} for
details on the SR expansion method)
\begin{equation}
\nJ^{MEC}_{SR} = \frac{1}{\sqrt{1+\tau}}\nJ^{MEC}_{NR} \, ,
\end{equation}
where $\nJ^{MEC}_{NR}$ is the traditional non-relativistic MEC operator.

\begin{figure}[tb]
\begin{center}
\leavevmode
\def\epsfsize#1#2{1.0#1}
\epsfbox[100 580 500 700]{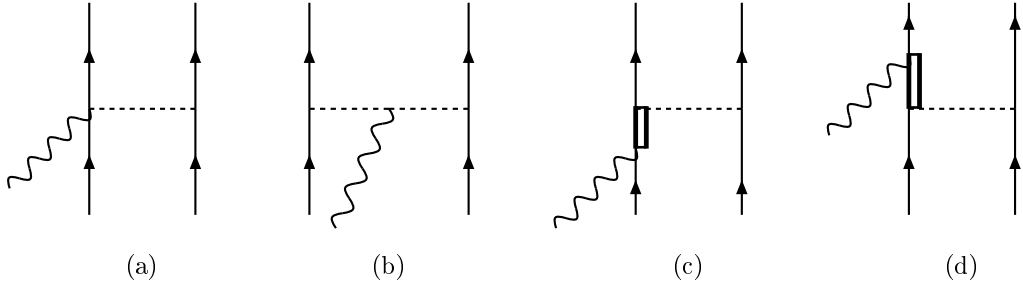}
\end{center}
\caption{MEC diagrams contributing to the two-body current of this work:
seagull (a), pion in flight (b) and $\Delta$ (c,d) currents are 
considered including relativistic corrections.}
\end{figure}

The expressions for the reduced matrix elements of the OB and MEC
multipole operators (\ref{Csigma},\ref{Esigma},\ref{Msigma}) in the
shell model are given in Refs.~\cite{Ama93,Ama94,Ama96b} except for
the relativistic correction factors appearing within the SR operators.
The somewhat complex structure displayed by these multipoles makes it
not possible to predict the relative importance of each contribution separately
without explicit numerical evaluation, even in the case of the OB
current.
 
Although in this work we perform a DWIA analysis of the response
functions, we may take advantage of the significant simplifications
introduced within the plane wave impulse approximation (PWIA), where
analytical expressions for the response functions can be
obtained~\cite{Mar02a,Mar02b}. First, for intermediate to high values
of $q$, the PWIA approach is expected to provide reasonable results,
thus the analytical PWIA expressions allow us to estimate the
contributions of the different pieces of the currents to the polarized
response functions. 
Second, since the PWIA
results should be recovered using the present multipole expansion in
the limit of no FSI, the comparison between our calculation and the
analytical PWIA responses makes it possible to fix the number of
multipoles needed to get convergence.

Hence, within PWIA, the matrix element of the OB current is written as
\begin{equation}
\langle \np'\ns,B|J^\mu(\nq)|A\rangle
=\sum_{\beta'\beta}
{\cal D}^*_{\beta'\frac12}(\ns) 
J^{\mu}(\np',\np)_{\beta'\beta}
\langle B|a_{\np,\beta}|A\rangle \, ,
\end{equation}
where $a_{\np,\beta}$ is the annihilation operator corresponding to a
particle with momentum $\np$ and spin projection $\beta$ referred to
the quantization axis.  Inserting this expression into the hadronic
tensor (\ref{hadronic-tensor}), and following the procedure described
in~\cite{Ama96b}, we obtain
\begin{equation}\label{factorized}
W^{\mu\nu} = \frac12 m_Np'  w^{\mu\nu}(\np',\np,\ns)M^S(\np) \, ,
\end{equation}
where we have defined the polarized single-nucleon tensor
\begin{equation}
w^{\mu\nu}(\np',\np,\ns) = 
\sum_{\alpha\alpha'\beta'}
{\cal D}^*_{\beta'\frac12}(\ns) 
J^{\nu}(\np',\np)_{\beta'\alpha}
J^{\mu}(\np',\np)_{\alpha'\alpha}^*
{\cal D}_{\alpha'\frac12}(\ns) \, .
\end{equation}
In the case of interest here, a closed-shell nucleus, the scalar momentum
distribution $M^S(\np)$ 
for nucleon knock-out from a shell $nlj$ is given by 
\begin{equation}
M^S(\np)=\frac{2j+1}{4\pi}\widetilde{R}^2(p)
\end{equation}
with $\widetilde{R}(p)$  the radial wave function of the hole 
in momentum space.

Using the current matrix elements (\ref{rho},\ref{Jx},\ref{Jy}) one
can compute in the factorized approximation (\ref{factorized})
the response functions (\ref{rlrt}--\ref{rt'rtl'}). From these results
the PWIA expressions for the reduced response functions can be identified.
Expressions for the unpolarized responses in PWIA were given in~\cite{Ama96b,Ama98b}. 
In the case of the polarized responses, from the total of eighteen, only five survive in PWIA. 
These are given by
\begin{eqnarray}
W^{T'}_{11}
&=& \frac{m_Np'}{4\pi} 2J_cJ_m\chi M^S(\np)
\label{WT'11-pw}
\\
W^{T'}_{10}
&=& \frac{m_Np'}{4\pi} 2J_m^2 M^S(\np)
\label{WT'10-pw}
\\
\widetilde{W}^{TL'}_{11}
&=& \frac{m_Np'}{4\pi} 2\sqrt{2} (\rho_cJ_m-\rho_{so}J_c\chi^2) M^S(\np)
\\
W^{TL'}_{11}
&=& \frac{m_Np'}{4\pi} 2\sqrt{2}\rho_cJ_m M^S(\np)
\\
W^{TL'}_{10}
&=& \frac{m_Np'}{4\pi} 2\sqrt{2}\rho_{so}J_m\chi M^S(\np) \, ,
\label{WTL'10-pw}
\end{eqnarray}
where we have used the factors introduced in
(\ref{rho-factors},\ref{J-factors}). Notice that all the $L,T,TL$ and
$TT$-type polarized responses are zero in this approximation.


\section{Results}


In this section we present results for selected recoil nucleon
polarization observables corresponding to proton knockout from the
$p_{1/2}$ and $p_{3/2}$ shells in $^{16}$O. In particular, we restrict
ourselves to the analysis of all the polarized response functions,
including the fifth response $W^{TL'}_0$ that does not depend on the
nucleon polarization and only enters when the initial electron beam is
polarized, and the transferred polarization asymmetries
$P'_{l,t,n}$. The study of cross sections and induced polarizations
will be presented in a forthcoming publication~\cite{Kaz03}. Two different
kinematical situations corresponding to $(q,\omega)$-constant
kinematics (also referred as quasiperpendicular kinematics) have been
selected: i) $q=460$ MeV/c, $\omega=100$ MeV, and ii) $q=1$ GeV/c,
$\omega=450$ MeV. In both cases the value of the transfer energy
$\omega$ corresponds almost to the quasielastic peak.

In this work our main interest is focused on the role of the two-body
MEC operators upon the recoil nucleon polarization observables, trying
to identify kinematical conditions for which these effects can be
important; however, a brief excursion on the FSI effects is also
presented. All the calculations have been done within the formalism
described in the previous section, i.e., semirelativistic expressions
for the one- and two-body current operators and a multipole expansion
method have been used. The number of multipoles needed has been fixed
by comparing the DWIA results, in the particular case of no FSI, with
the exact factorized PWIA
responses~(\ref{WT'11-pw}--\ref{WTL'10-pw}). Convergence in the
multipole analysis is obtained with $J_{max}=30$ for $q=460$ MeV/c,
and $J_{max}=35$ for $q=1000$ MeV/c.  Finally, in all of the results
which follow, the kinematics of the ejected nucleon is treated
exactly.

\subsection{Polarized response functions}

Here we analyze the thirteen responses defined
in~(\ref{WKn}--\ref{WKt}) which arise from the ejected nucleon
polarization, plus the ``fifth'' response function
$W^{TL'}_o$. Results are displayed in Figs.~3--12.  A similar analysis
for the unpolarized responses $L$, $T$, $TL$ and $TT$ has been
performed recently in~\cite{Ama03b}.

\begin{figure}[p]
\begin{center}
\leavevmode
\def\epsfsize#1#2{0.9#1}
\epsfbox[100 260 500 780]{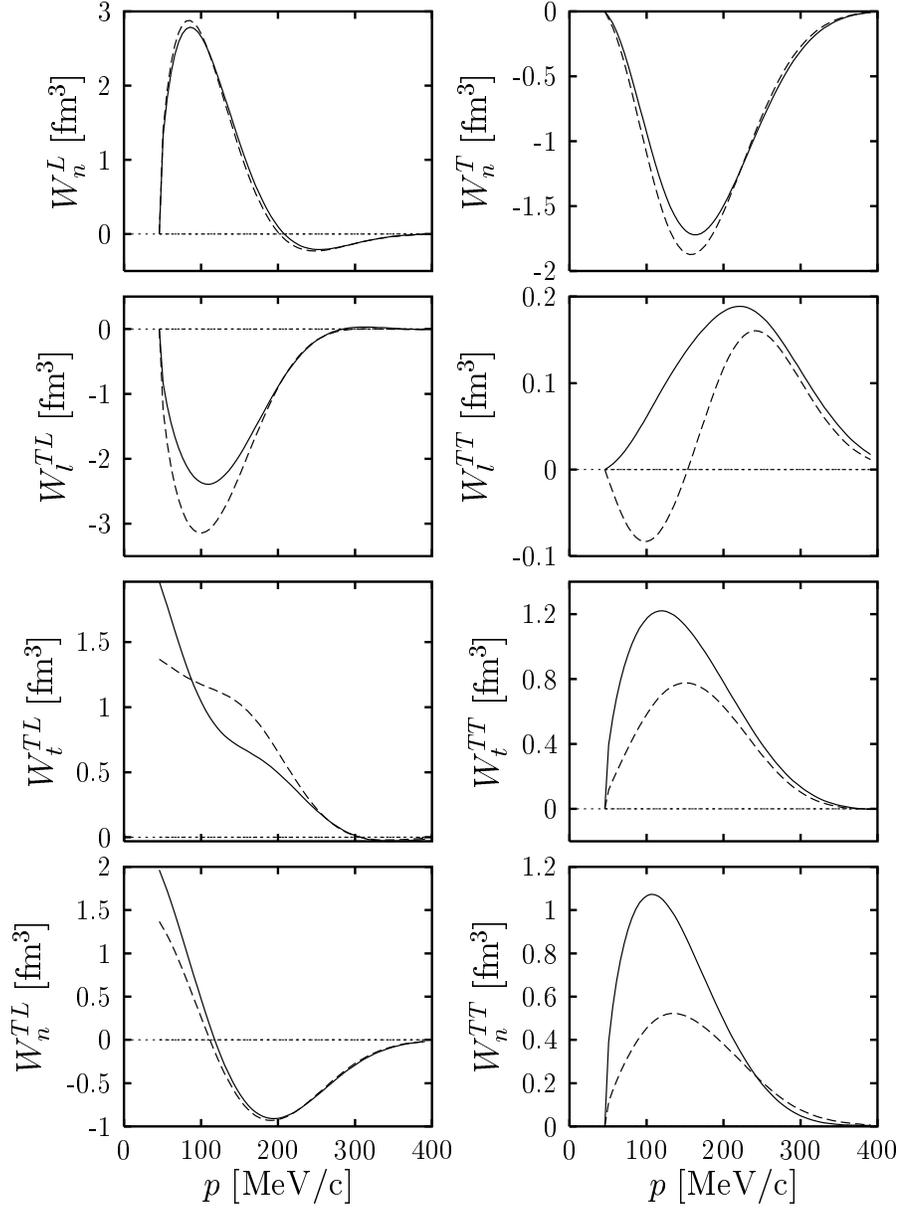}
\end{center}
\caption{\label{fig-v1} Induced polarized response functions ($L$, $T$, $TL$
and $TT$--type responses) for proton knock-out from the $1p_{1/2}$ shell in
$^{16}$O, with momentum transfer $q=460$ MeV/c and energy transfer
$\omega=100$ MeV. Solid lines are the DWIA results using the optical
potential of Comfort and Karp; dashed lines have been computed with
the Schwandt optical potential.}
\end{figure}

\begin{figure}[hp]
\begin{center}
\leavevmode
\def\epsfsize#1#2{0.9#1}
\epsfbox[100 380 500 780]{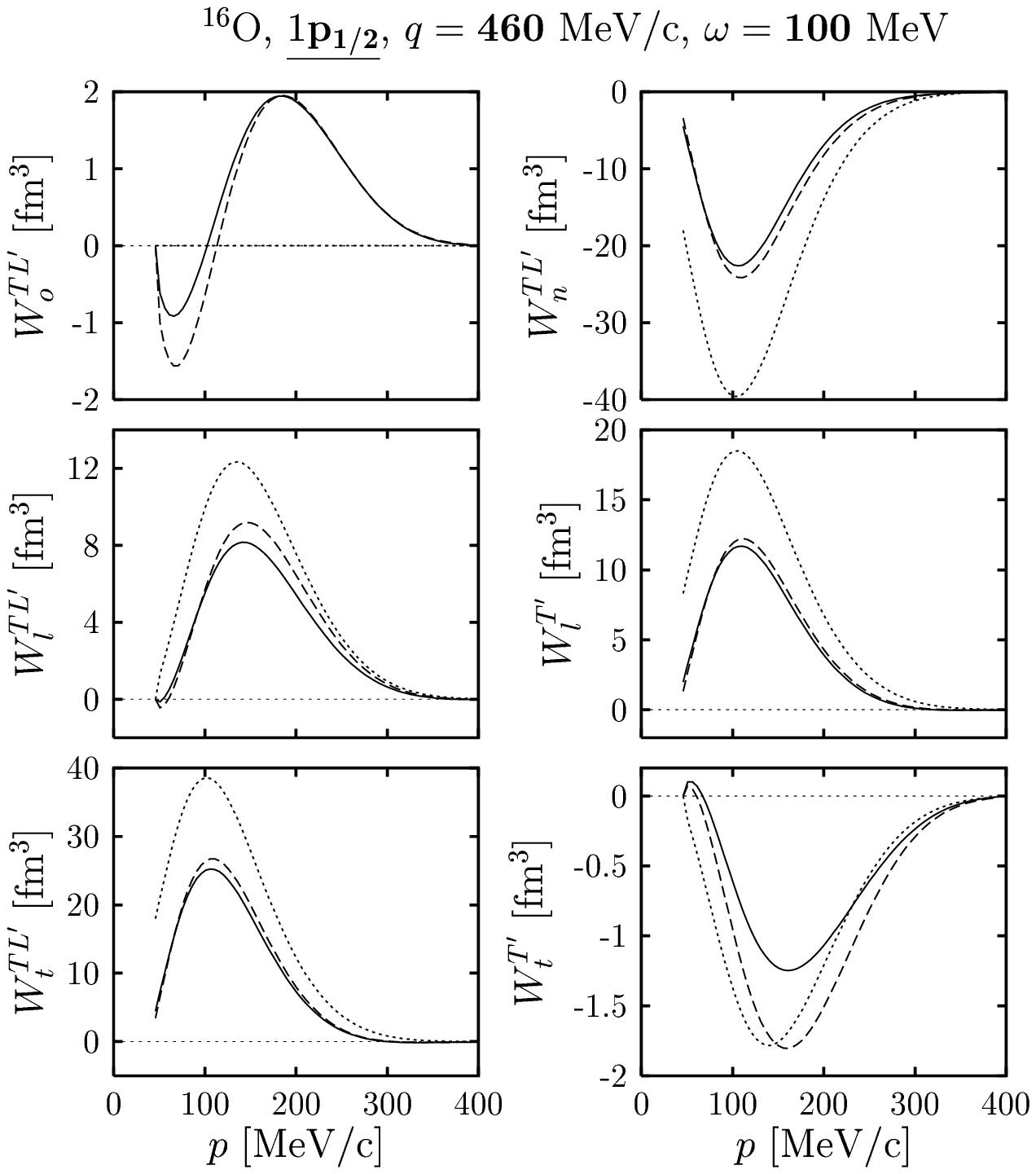}
\end{center}
\caption{\label{fig-v2} The same as fig.~3 for the fifth and
transferred polarized response functions ($T'$ and $TL'$--type
responses). With dotted lines we show also for reference the PWIA
results.}
\end{figure}

\subsubsection{Effects of FSI}

We start our discussion with the effects of FSI on the polarized
responses. A study of the dependence of the unpolarized responses on
the particular FSI model was already presented in~\cite{Ama99}. In
Fig.~3 we show the eight induced polarized responses for proton
knock-out from the $1p_{1/2}$ shell in $^{16}$O as function of the
missing momentum $p$. Kinematics corresponds to $q=460$ MeV/c and
$\omega=100$ MeV.  The five transferred polarized responses plus the
fifth one ($T'$ and $TL'$ types) are displayed in Fig.~4. Similar
results are obtained for the $1p_{3/2}$ shell and thus they are not
shown here.  In all of these results we use bound wave functions
obtained as solutions of the Shr\"odinger equation with a Woods-Saxon
potential, with parameters taken from ref.~\cite{Ama96a}. For the
final states we use solutions for two different optical
potentials. Solid lines correspond to calculations performed with the
Comfort and Karp potential~\cite{Com80}, which was originally fitted
to elastic proton scattering from $^{12}$C for energies below 183
MeV. We have extended it to $^{16}$O by introducing a dependence
$A^{1/3}$ in the radius parameters. The results shown with dashed
lines have been computed with the Schwandt potential~\cite{Sch82},
which also has been extrapolated here for $^{16}$O since it was
originally fitted to higher mass nuclei.

The induced polarized $L,T,TL,TT$ and the fifth response functions,
which are zero in absence of FSI, are expected to be highly sensitive
to the details of the particular optical potential considered, and in
particular, to the spin-orbit term in the potential. In this sense
notice the significant difference introduced by both potentials in the
case of the polarized $TL$ and $TT$ responses (fig.~\ref{fig-v1}),
while the FSI discrepancy gets smaller for the fifth response function
(fig.~\ref{fig-v2}) and is considerably reduced for $W^L_n$ and
$W^T_n$.

The five transferred polarized responses which survive in
PWIA~(\ref{WT'11-pw}--\ref{WTL'10-pw}), depend less on the details of
the potential, being mostly affected by the central imaginary part of
it. As known, these responses enter in the case in which also the
initial electron is polarized, and they contribute to the transferred
nucleon polarization asymmetry.  We observe (fig.~\ref{fig-v2}) that
both potentials lead to close results, differing by less than 10\% for
the dominant responses $W^{TL'}_n$, $W^{T'}_l$ and $W^{TL'}_l$, while
the largest differences are shown for $W^{T'}_t$, which is however very small. 
Similar results are found for the
$p_{3/2}$ shell. The sensitivity shown by some polarized responses to
the details of the potential, makes these observables of special
interest  to disentangle between the different
models of FSI that can fit reasonably well the unpolarized cross
sections.

\begin{figure}[p]
\begin{center}
\leavevmode
\def\epsfsize#1#2{0.9#1}
\epsfbox[100 260 500 780]{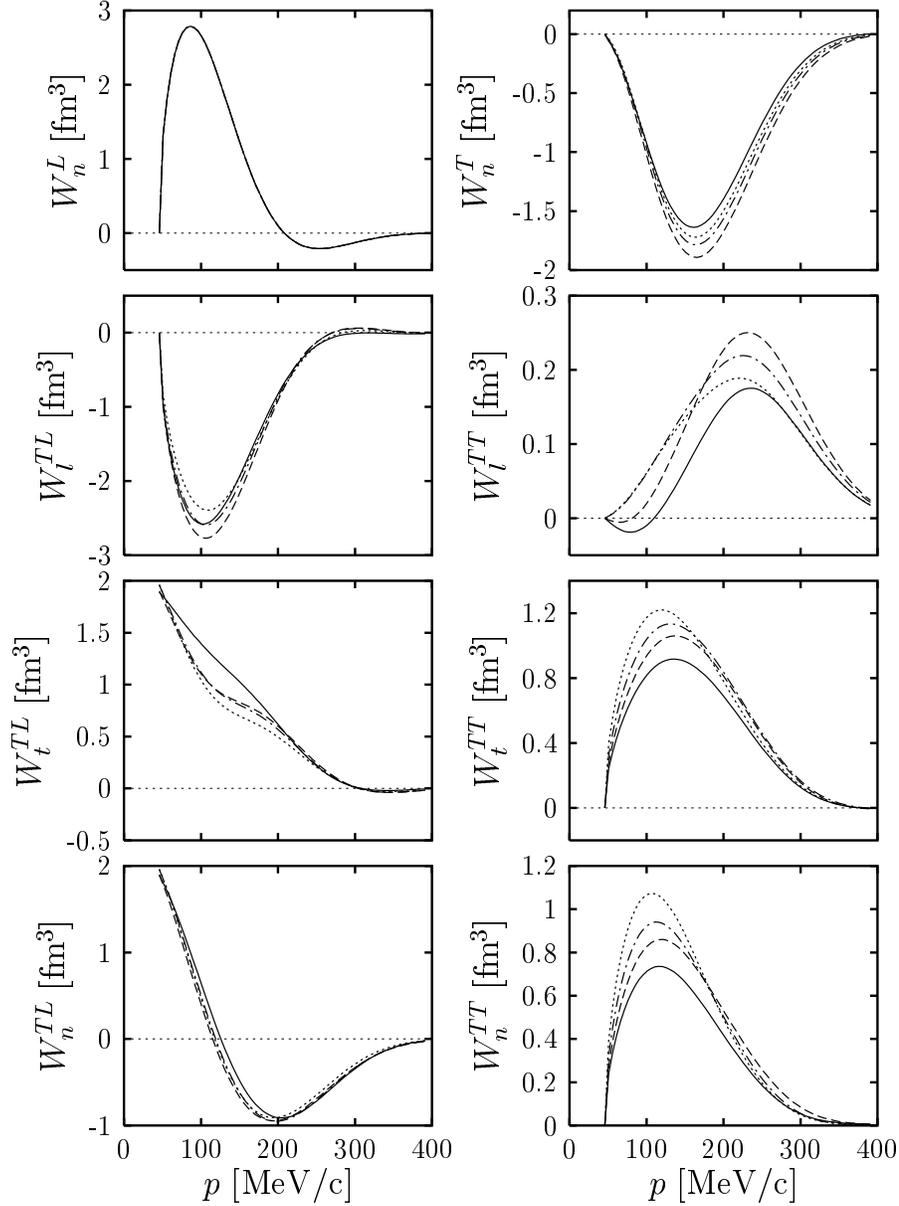}
\end{center}
\caption{\label{fig-mec1}
Induced polarized response functions 
for proton knock-out from the $1p_{1/2}$ shell in $^{16}$O, 
with momentum transfer $q=460$ MeV/c and
energy transfer $\omega=100$ MeV. 
Dotted lines are the DWIA results using only the OB current operator;
dashed lines include in addition the seagull current (OB+S); 
dot-dashed include the seagull plus pionic currents (OB+S+P);
finally solid lines represent the total 
result (OB+MEC) including also the $\Delta$ current.
}
\end{figure}

\begin{figure}[hp]
\begin{center}
\leavevmode
\def\epsfsize#1#2{0.9#1}
\epsfbox[100 380 500 780]{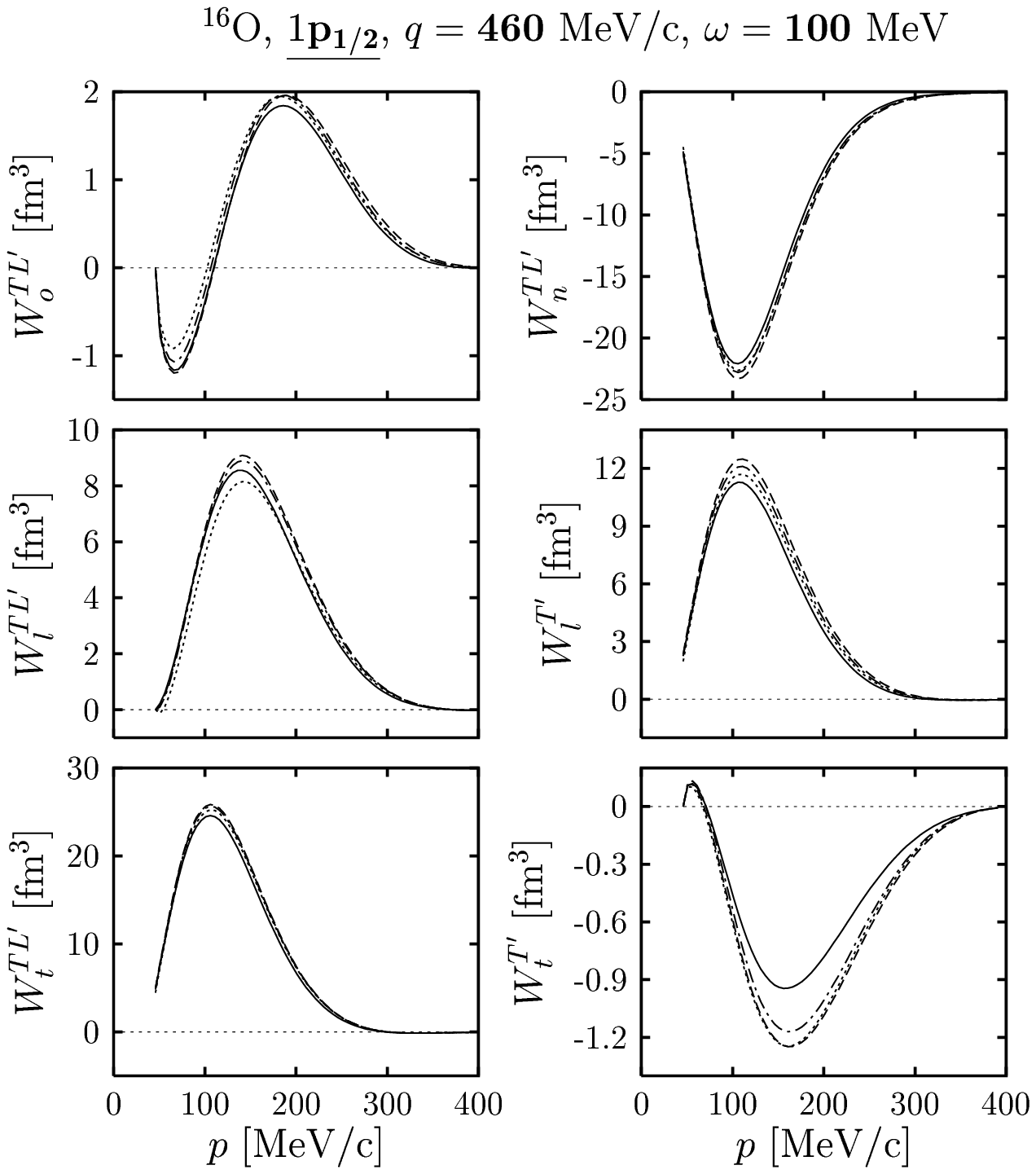}
\end{center}
\caption{\label{fig-mec2}
Fifth and transferred polarized response functions
for $^{16}$O. Kinematics corresponds to momentum transfer $q=460$ MeV/c and
energy transfer $\omega=100$ MeV. 
The meaning of the lines is the same as in Figure 5.
}
\end{figure}

\subsubsection{Effects of MEC}

\begin{figure}[hp]
\begin{center}
\leavevmode
\def\epsfsize#1#2{0.9#1}
\epsfbox[100 260 500 780]{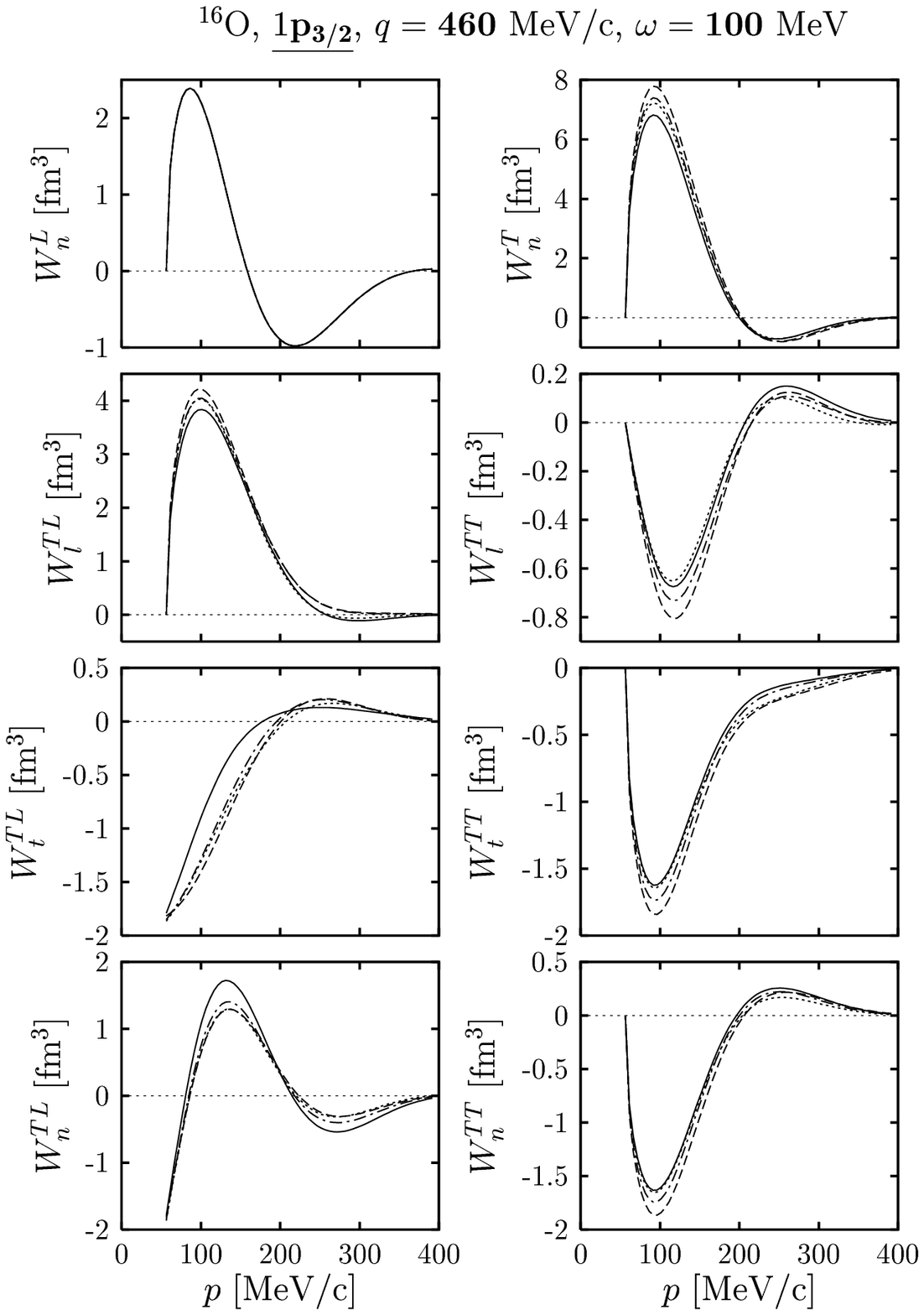}
\end{center}
\caption{
\label{fig-mec3}
The same as Fig.~5 for proton knock-out from the $1p_{3/2}$ shell.
}
\end{figure}

\begin{figure}[hp]
\begin{center}
\leavevmode
\def\epsfsize#1#2{0.9#1}
\epsfbox[100 380 500 780]{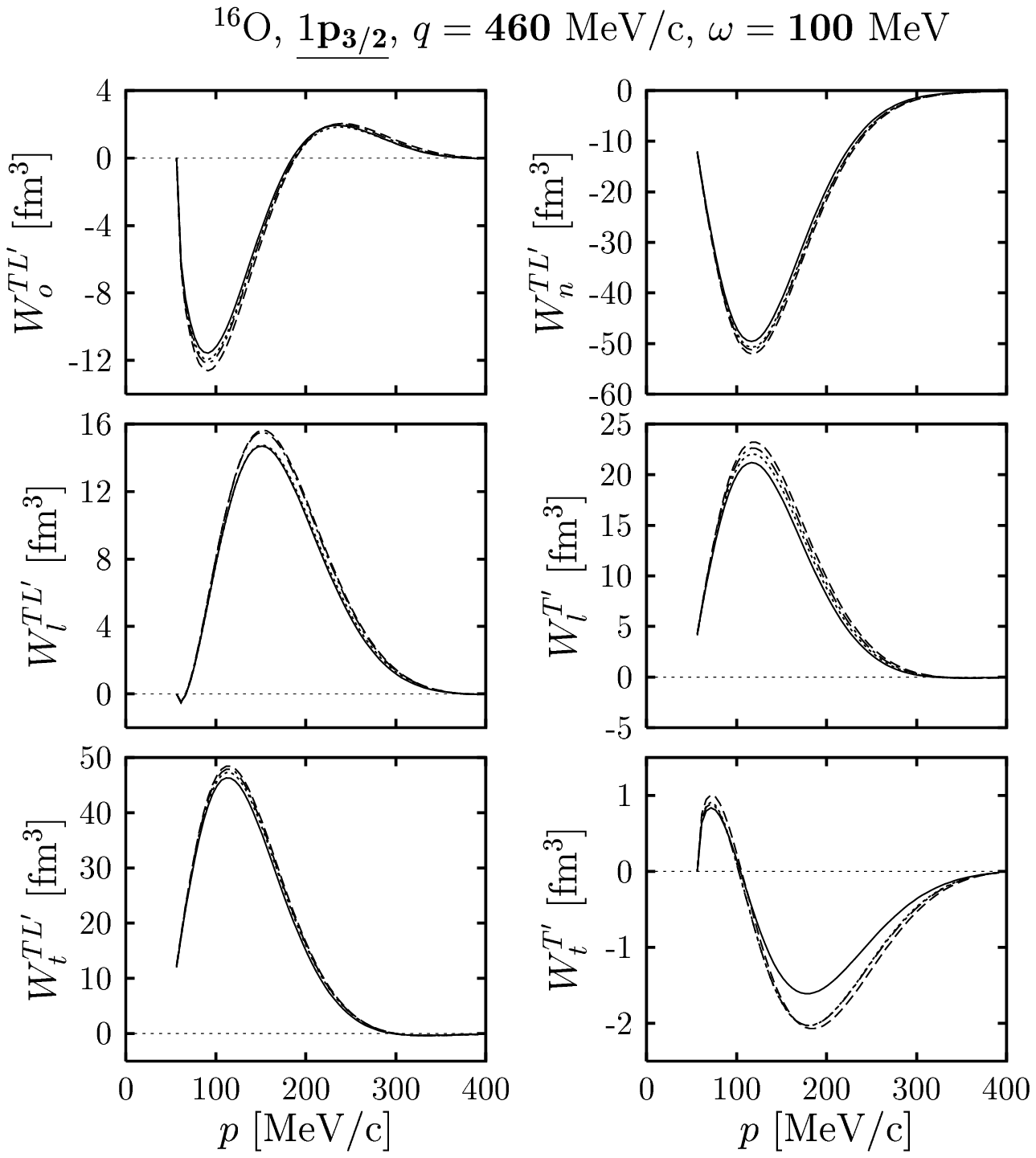}
\end{center}
\caption{
\label{fig-mec4}
The same as Fig.~6 for proton knock-out from the $1p_{3/2}$ shell.
}
\end{figure}

\begin{figure}[hp]
\begin{center}
\leavevmode
\def\epsfsize#1#2{0.9#1}
\epsfbox[100 260 500 780]{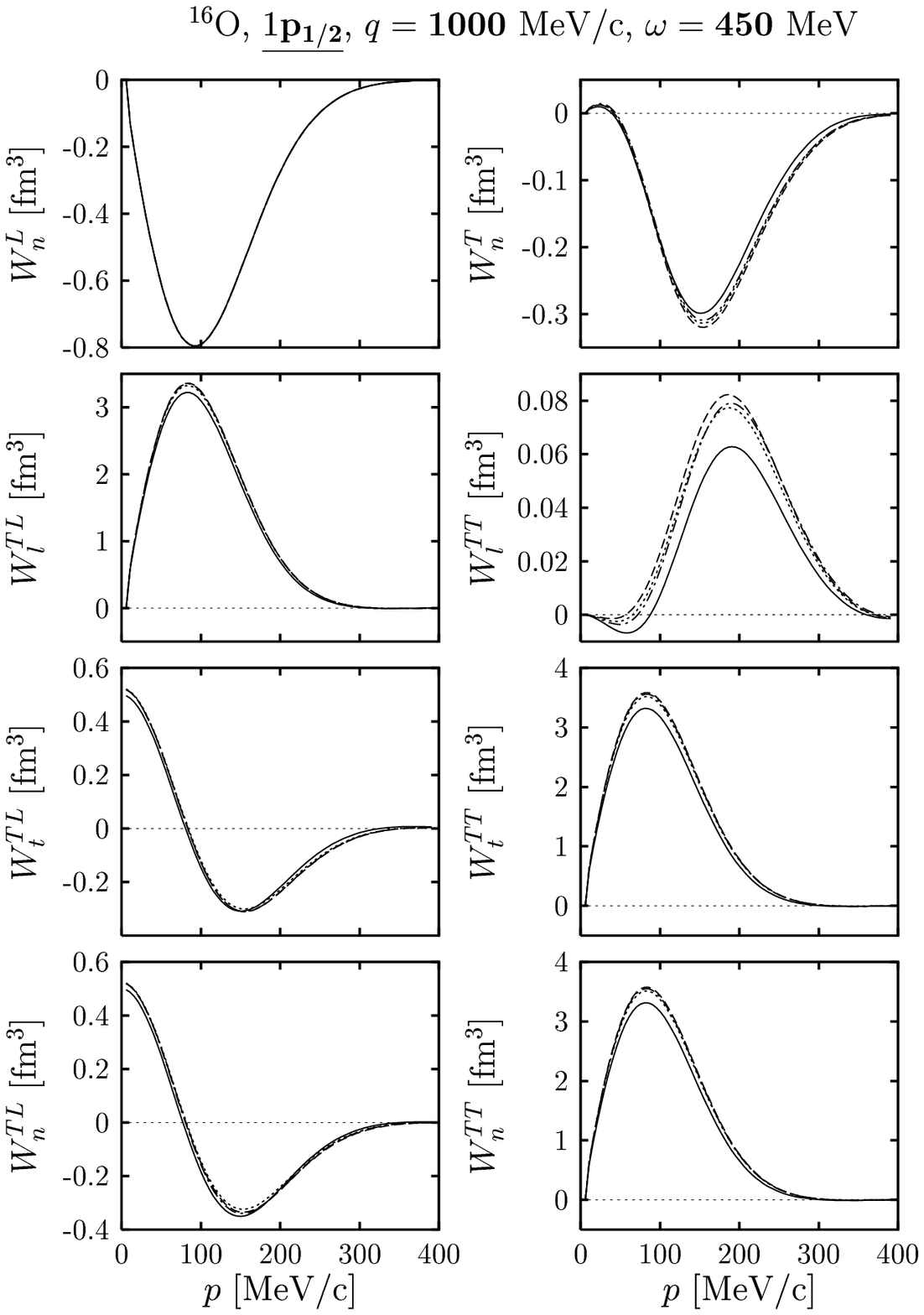}
\end{center}
\caption{
\label{fig-mec5}
The same as Fig.~5 for $q=1000$ MeV/c and $\omega=450$ MeV.
}
\end{figure}

\begin{figure}[hp]
\begin{center}
\leavevmode
\def\epsfsize#1#2{0.9#1}
\epsfbox[100 380 500 780]{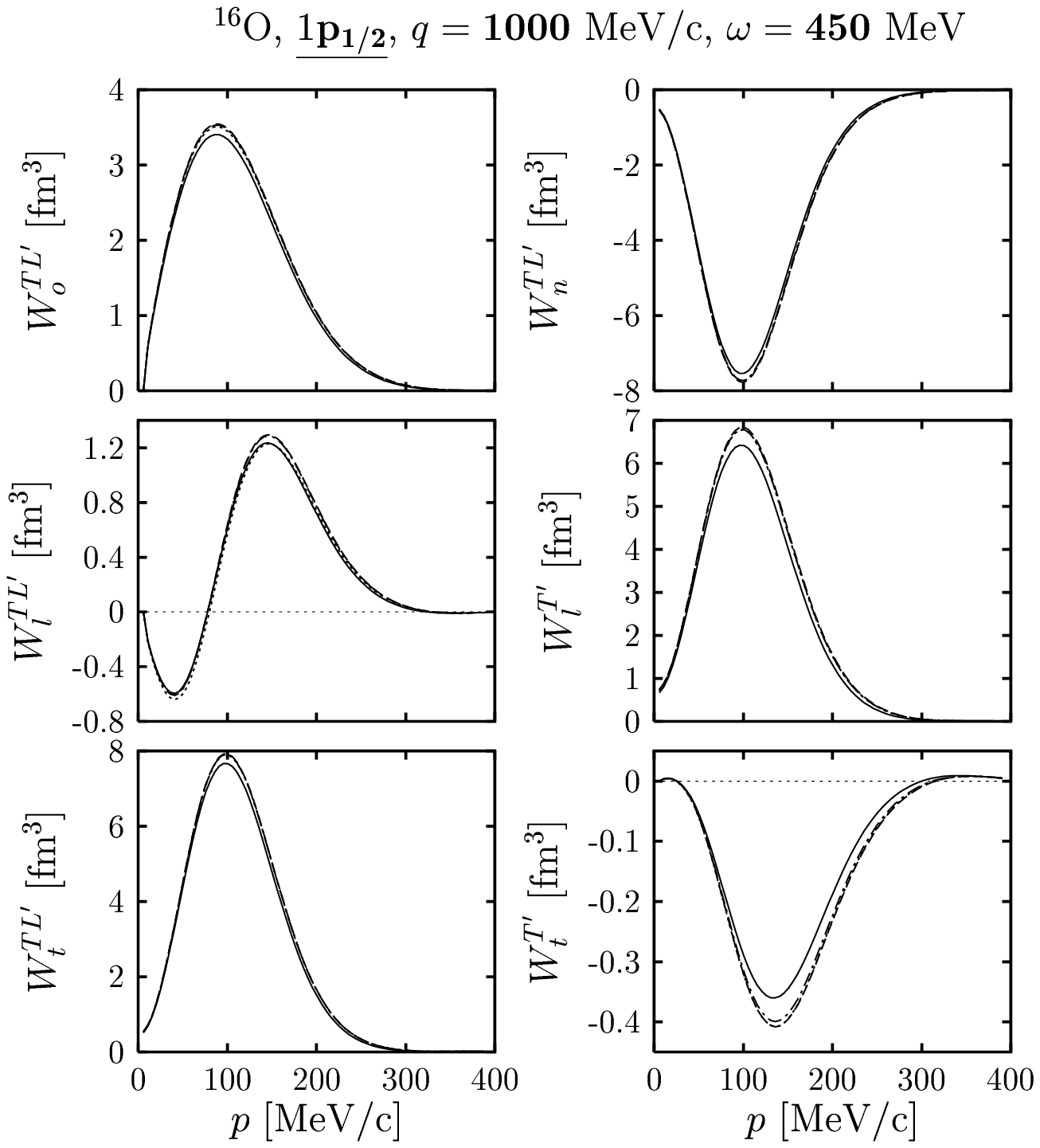}
\end{center}
\caption{
\label{fig-mec6}
The same as Fig.~6 for $q=1000$ MeV/c and $\omega=450$ MeV.
}
\end{figure}

\begin{figure}[hp]
\begin{center}
\leavevmode
\def\epsfsize#1#2{0.9#1}
\epsfbox[100 260 500 780]{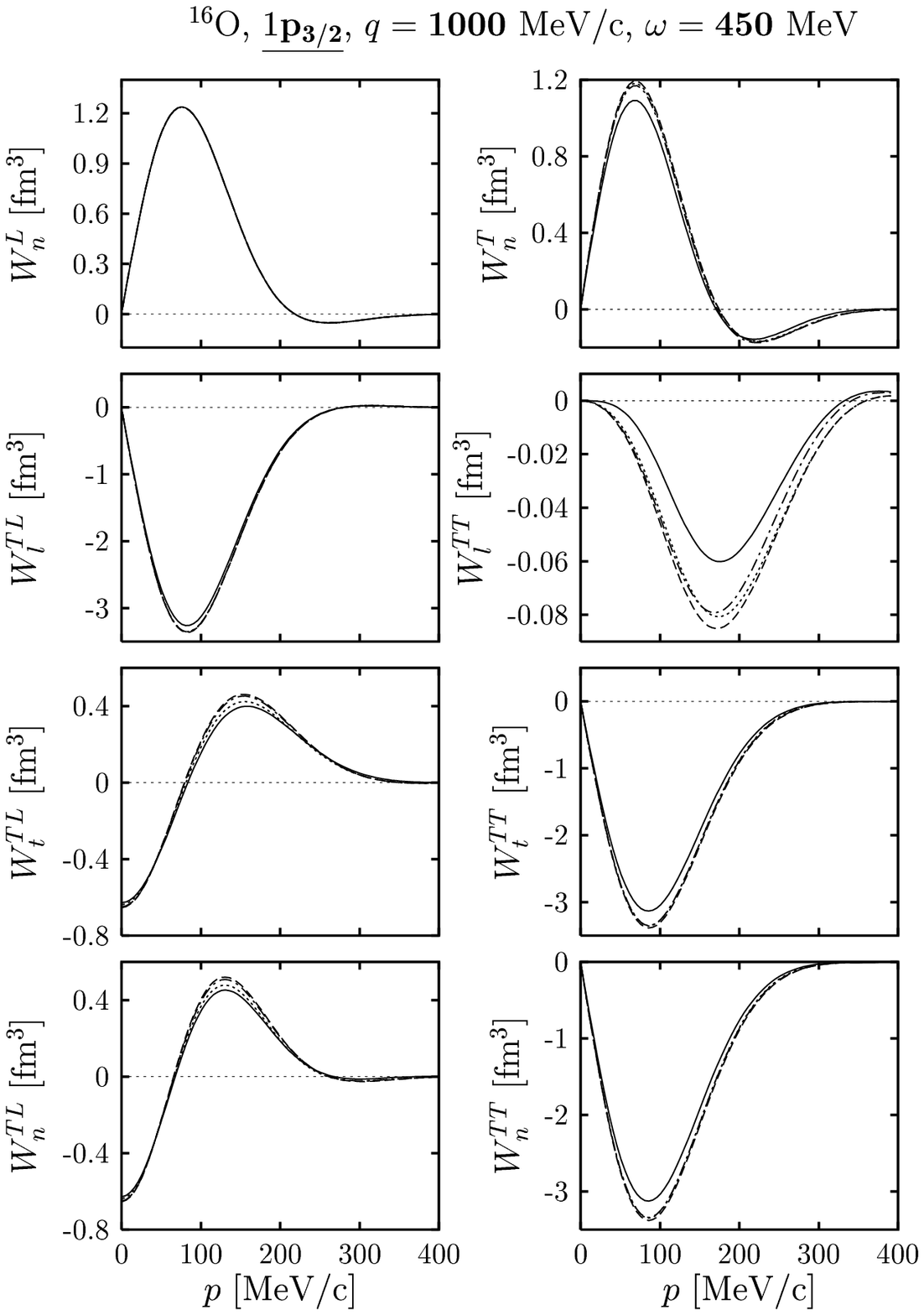}
\end{center}
\caption{
\label{fig-mec7}
The same as Fig.~5 for the $p_{3/2}$ shell, $q=1000$ GeV/c, and
$\omega=450$ MeV.  }
\end{figure}

\begin{figure}[hp]
\begin{center}
\leavevmode
\def\epsfsize#1#2{0.9#1}
\epsfbox[100 380 500 780]{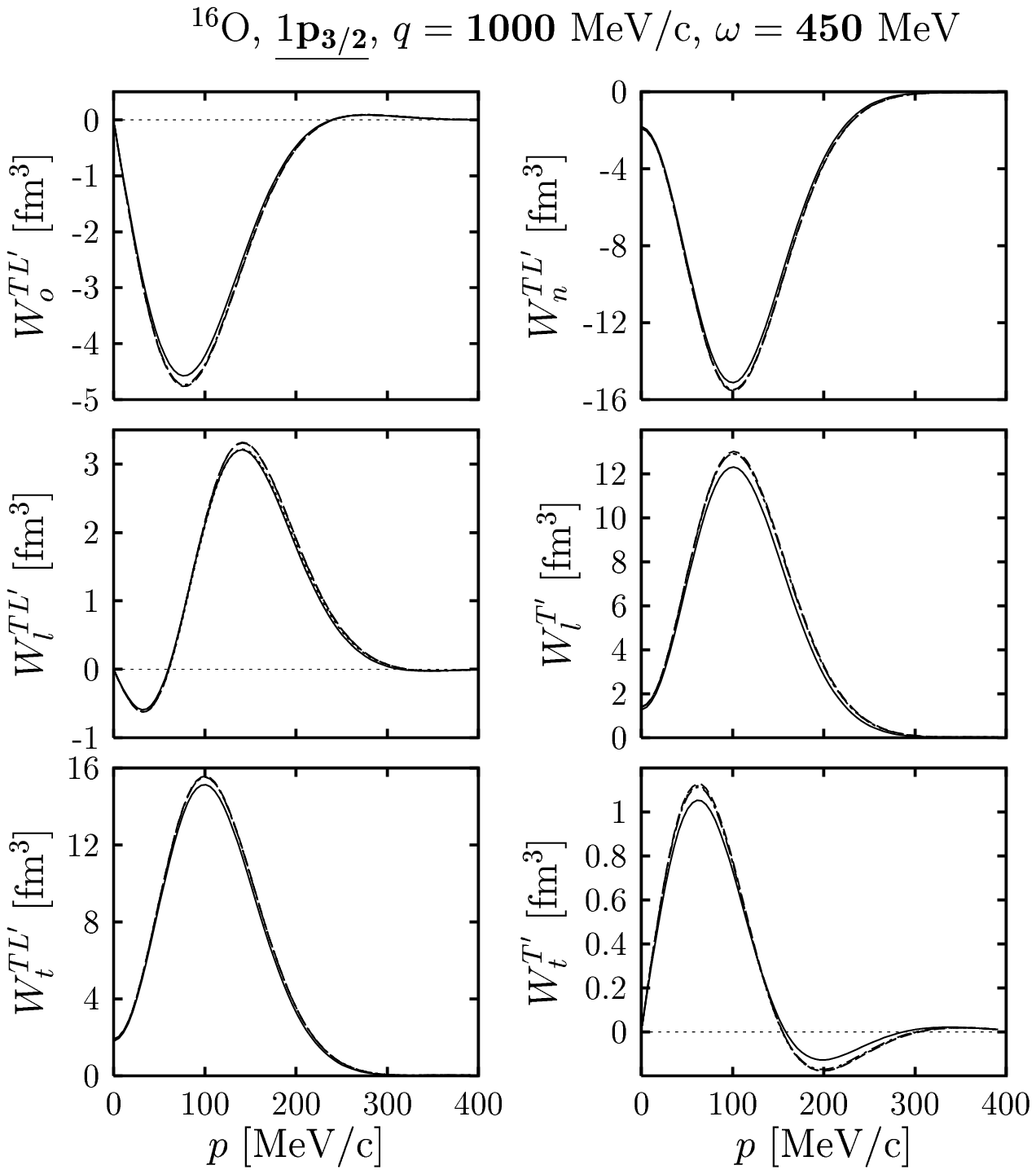}
\end{center}
\caption{
\label{fig-mec8}
The same as Fig.~6 for the $p_{3/2}$ shell, 
$q=1000$ MeV/c and $\omega=450$ MeV.
}
\end{figure}

The impact of MEC on the recoil nucleon polarized responses is shown
in Figs.~\ref{fig-mec1}--\ref{fig-mec8}.  In each panel we compare the
distorted wave responses evaluated by using the OB current only
(dotted line) with the results obtained when including also the
two-body MEC operators considered in fig.~2, namely, the seagull or
contact (OB+S) current (dashed lines), the contact and pion-in-flight
(OB+S+P) currents (dot-dashed lines), and finally, including also the
$\Delta$ current (solid lines), denoted as (OB+MEC). Results in
figs.~5--8 correspond to kinematics $q=460$ MeV/c, $\omega=100$ MeV
(kinematics I), whereas in figs.~9--12 we present the responses
evaluated at $q=1$ GeV/c, $\omega=450$ MeV (kinematics II). For both
kinematics proton knock-out from the $p_{1/2}$ (figs.~5,6 and 9,10)
and $p_{3/2}$ (figs.~7,8 and 11,12) have been considered.

Let us discuss first the results for kinematics I (figs.~5-8). Here we
observe that the global sign of the polarized $T$, $TL$ and $TT$
responses changes when comparing the $p_{1/2}$ (fig.~\ref{fig-mec1})
and $p_{3/2}$ (fig.~\ref{fig-mec3}) shells.  The same occurs for
kinematics II. Concerning MEC effects, the various polarized responses
display different sensitivities to the two-body component of the
nuclear current. Apart from the pure longitudinal response $W^L_n$,
which shows no dependence on MEC because the ``semi-relativistic'' MEC
expressions only include the leading transverse components, the role
of MEC on $W^T_n$ is shown to be similar to the one found for the
unpolarized $T$-response in~\cite{Ama03b}: the enhancement (in
absolute value) produced by the S current is partially cancelled by
the reduction introduced by the P current; the $\Delta$ current gives
rise to an additional reduction, leading to a global decrease of the
$W^T_n$ response of the order of $\sim 10\%$ at the maximum.  This
effect being similar for both shells (figs.~\ref{fig-mec1} and
\ref{fig-mec3}).

Larger MEC effects are found for some of the induced polarized $TL$
responses, particularly for $W^{TL}_t$ where the $\Delta$ current
produces a very significant modification of the response, changing
even its shape in the region close to $p\sim 100$ MeV/c.
Note that,
although the global effect introduced by the $\Delta$ in this response
is similar for both shells, in the case of the $p_{1/2}$ there is a
large increase, whereas for $p_{3/2}$ the response is significantly
reduced in absolute value. It is also interesting to point out that
the $\Delta$ current plays also the most important role for the
$W^{TL}_n$ response, this being clearly shown in the case of the
$p_{3/2}$ shell.

The role of MEC on the three polarized $TT$ responses shows a very
different behaviour for the two shells considered. In the case of the
$p_{1/2}$ (fig.~\ref{fig-mec1}), the global effect of MEC is a very
significant reduction of the responses, particularly for $W^{TT}_t$
($\sim 20\%$) and $W^{TT}_n$ ($\sim 30\%$), being the separate
contributions of the S, P and $\Delta$ currents of rather similar
importance. 
Note that the contributions introduced by the S and P
currents have opposite signs for the $p_{1/2}$ and $p_{3/2}$ shells.
As a consequence, for the $p_{3/2}$ shell (fig.\ref{fig-mec3}) the large
enhancement (in absolute value) produced by the S current is almost
cancelled exactly by the contributions of the P and $\Delta$
currents, so the net MEC effect is almost negligible for the three $TT$
responses.

The transferred polarized responses ($T'$ and $TL'$-type responses)
are shown in Figs.~\ref{fig-mec2} and \ref{fig-mec4}. From these
results we find in general a small effect of MEC, less than $\sim
5\%$. An exception is $W^{T'}_t$ where the role of $\Delta$ gives rise
to an important reduction of the response; however notice that
$W^{T'}_t$ is very small, of the order of $\sim 10\%$ compared with
$W^{T'}_l$ and $W^{TL'}_t$, and hence difficult to measure.  The
anomalous smallness of the response $W^{T'}_t$ was already discussed
in detail in~\cite{Mar02b} within the context of the PWIA and
different non-relativistic reduction schemes. This result can be also
understood within the multipole analysis performed in this work by
taking into account the general relations given in
eqs.~(\ref{WKl},\ref{WKt}) and the explicit expressions obtained for
the multipole functions in PWIA~(\ref{WT'11-pw},\ref{WT'10-pw}).
Since we are close to the
quasielastic-peak, the angle $\theta'$ is close to zero for moderate
missing momentum, so the biggest contribution comes from the factor
multiplied by $\cos\theta'$ in eqs.~(\ref{WKl},\ref{WKt}).  
This factor is $W^K_{10}=O(1)$ in the case of
the $l$-responses, and $W^K_{11}=O(\chi)$ for the $t$-response. 
Precise values of the nucleon form factors and kinematical variables
can be introduced in these equations to verify the exact relation
between the $l$ and $t$ components in PWIA, which is not very
different from the distorted-wave results of figs.~\ref{fig-mec2}
and~\ref{fig-mec4}.

Results for higher momentum and energy transfer, $q=1000$ MeV/c and
$\omega=450$ MeV (kinematics II), are shown in
Figs.~\ref{fig-mec5}--\ref{fig-mec8} for the two $p$ shells in
$^{16}$O. This kinematics corresponds to the experimental setting
of~\cite{Gao00,Mal00} where $Q^2=-0.8$~(GeV/c)$^2$. Obviously in this
case relativity is expected to play a more important role and in fact,
studies within the relativistic distorted wave impulse approximation
(RDWIA)~\cite{Udi99} have proved the importance of these effects. The
present SR model, although lacking some of the relativistic
ingredients inherent in the RDWIA, incorporates exact relativistic
kinematics for the ejected nucleon, a SR expansion of the current
which can be used for high $q$ values, and finally, the use of the
Schr\"odinger-equivalent form of the S-V Dirac global optical
potential of~\cite{Coo93}, including the Darwing term in the wave
function. The validity of the expansion procedure used in the SR model
was tested in~\cite{Udi99} where unpolarized observables evaluated
within the SR approach were compared with a RDWIA calculation for this
kinematics.

The discussion of the results presented in
figs.~\ref{fig-mec5}-\ref{fig-mec8} follows similar trends to the ones
already presented for kinematics I, so here we simply summarize those
aspects which can be of more relevance. As shown in
figs.~\ref{fig-mec5}-\ref{fig-mec8}, the general effect introduced by
MEC is a global reduction of the responses (in absolute value) whose
magnitude depends on the specific response, being of the order of a
few percent for  $W^{TL}_{l,t,n}$ and $W^{TL'}_{0,l,t,n}$,
larger for $W^{T}_{n}$, $W^{TT}_{t,n}$ and $W^{T'}_{l,t}$
(particularly because of the $\Delta$-contribution) and the largest
for $W^{TT}_l$, where the reduction (basically due to $\Delta$)
is about $\sim 20-25\%$.
Note however that the response $W^{TT}_l$ is the smallest one and so
hardly measurable.

The dependence of MEC effects on the momentum transfer shown in the
results of figs.~5--12 is consistent with the findings of
ref.~\cite{Ama03b} for the unpolarized responses. In general the
importance of MEC decreases with $q$.  
This is in accord with the results for the $T$ response in the $1p-1h$ channel
in the case of quasielastic inclusive $(e,e')$ reactions~\cite{Ama02c,Ama03a}. This behaviour
can be roughly understood from the relativistic expressions for the particle-hole
transverse current matrix elements ${\nJ}_T(\np',\np)$ in Fermi gas~\cite{Ama02c,Ama03a},
and also from the traditional non relativistic expressions.
At the non relativistic level, the OB
current is dominated by the magnetization contribution which goes as $\sim q$. 
On the contrary, MEC present a much more complex dependence on $q$ and on 
the momenta of the two holes involved: the missing momentum $\np$ and an
intermediate momentum $\nk$ which should be integrated. 
Moreover, MEC also contain pion propagators involving
inverse squared pion momenta. For high $q$, a crude estimation of the
(transverse) seagull and pion in flight currents is shown to behave as $\sim
q/(q^2+m_\pi^2)$, while the $\Delta$ current goes as $\sim
q^3/(q^2+m_\pi^2)$, hence the latter clearly dominates, which is in accord
with the results shown here. Once the $\pi N$ form factor, which becomes smaller when high
momenta are probed, is added to the two-body currents, we find the OB contribution to
dominate over the MEC. At the relativistic level the above dependences on $q$
change. In~\cite{Ama98a} it was demonstrated that if the form
factors are neglected, then the OB, seagull and pionic currents grow
asymptotically as $\sqrt{q}$. Thus the inclusion of $\pi N$ 
form factors is essential for the dominance of the OB current
This conclusion however applies to the response
functions only for low missing momentum, since for other observables
such as the $A_{TL}$ asymmetry \cite{Ama03b} and the polarization
asymmetries (see below) larger effects are found for high values of
$q$ and missing momentum.

\subsection{Transferred polarization asymmetries}

Apart from the response functions, other observables of special
interest are the nucleon polarization asymmetries introduced in
eqs.~(18-21).  These observables are given as ratios between polarized
and unpolarized responses, where one hopes to gain different insight
into the underlying physics from what is revealed through the
responses themselves. As already mentioned in the introduction and in
order to clarify the discussion, here we restrict ourselves to the
analysis of the transferred polarization asymmetries $P'_{l,t,n}$,
which only enter with polarized incident electrons and persist in
PWIA. Induced polarization ratios $P_{l,t,n}$ ---that do not depend on
the polarization of the incident electron and are zero within the
plane wave approach---, and total cross sections will be analyzed in a
forthcoming publication~\cite{Kaz03}.

Following the discussion presented for the responses, here we first
study the effects introduced by FSI and later on we focus on the role
of MEC.

\subsubsection{Effects of FSI}

\begin{figure}[p]
\begin{center}
\leavevmode
\def\epsfsize#1#2{0.8#1}
\epsfbox[100 170 500 775]{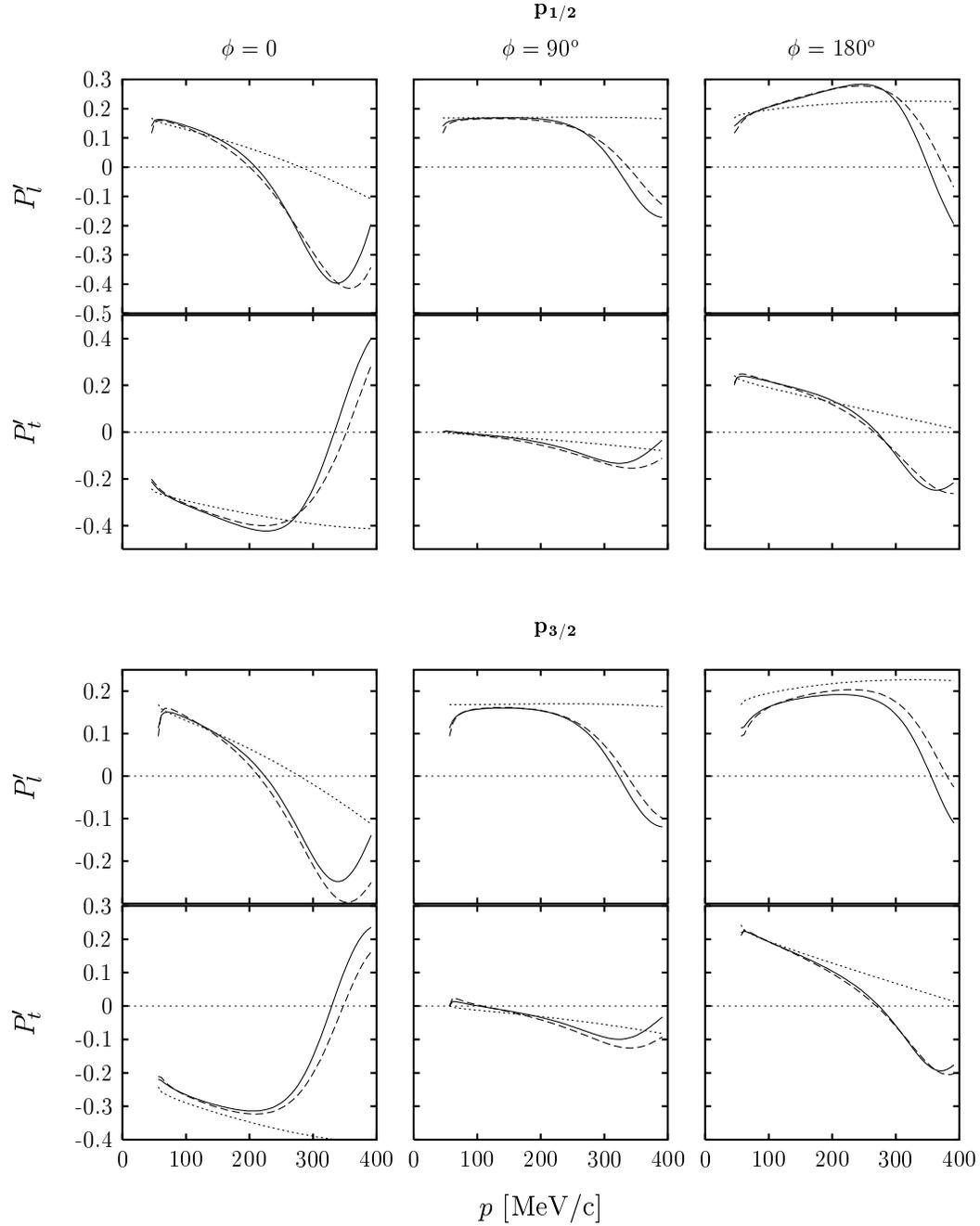}
\end{center}
\caption{ 
\label{fig-pol0}
Transferred polarization ratios $P'_l$ and $P'_t$ for proton
 knock-out from the $p$ shells in $^{16}$O, and $q=460$ MeV/c,
 $\omega=100$ MeV. The electron scattering angle is $\theta_e=30^{\rm
 o}$, and results are shown for three values of the proton azimuthal
 angle $\phi=0,90^{\rm o},180^{\rm o}$. Solid lines have been computed
 in DWIA with the Comfort and Karp potential, dashed lines with the
 Schwandt potential, and finally the dotted lines are the PWIA
 results.}
\end{figure}

\begin{figure}[tp]
\begin{center}
\leavevmode
\def\epsfsize#1#2{0.8#1}
\epsfbox[100 480 500 800]{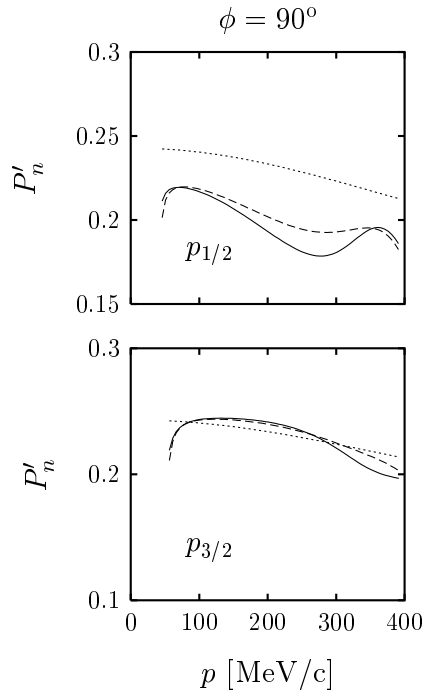}
\end{center}
\caption{ 
\label{fig-pol00}
The same as Fig.~\ref{fig-pol0} for the 
polarization transfer component $P'_n$ and $\phi=90^{\rm o}$.
}
\end{figure}

In Figs.~\ref{fig-pol0}--\ref{fig-pol00} we present the results
obtained for the transferred polarization asymmetries corresponding to
proton knockout from the $p_{1/2}$ and $p_{3/2}$ shells in
$^{16}$O. Kinematics has been selected as (I), i.e., $q=460$ MeV/c and
$\omega=100$ MeV. Results for kinematics (II) follow the same general
trends although FSI effects are in general less important because of
the higher momentum transfer involved. The
longitudinal $P'_l$ and transverse (sideways) $P'_t$ components are
shown in fig.~\ref{fig-pol0} for electron scattering angle fixed to
$\theta_e=30^{\rm o}$ (forward scattering) and three values of the
proton azimuthal angle $\phi=0,90^{\rm o}$ and 180$^{\rm o}$, while
the normal polarization $P'_n$ is displayed in fig.~\ref{fig-pol00}
for $\phi=90^{\rm o}$ (notice that $P'_n$ is zero for co-planar
kinematics). Although not shown here for brevity, we have also
explored the behaviour of the transferred polarization ratios at
backward scattering angle ($\theta_e=150^o$). As known, the purely
transverse responses dominate at backward angles, whereas all of the
kinematical factors that enter in the description of
$(\vec{e},e'\vec{p})$ reaction are of similar order at forward angles.
In~\cite{Mar02a} forward scattering angles were proved to enhance
significantly the sensitivity to dynamical relativistic
effects. Concerning FSI and MEC, the discussion of the results for
$\theta_e=150^{\rm o}$ follow similar trends to the ones presented here for
$\theta_e=30^{\rm o}$.

The PWIA calculation (dotted line) is compared with DWIA results using
the two optical potentials already presented in the previous section,
i.e., Comfort \& Karp (solid lines) and Schwandt (dashed lines).
First, notice the difference between PWIA and DWIA results. Within the
plane wave approach, the responses factorize and hence the
polarization ratios depend only on the single-nucleon responses, being
cancelled the whole dependence with the momentum distribution. This
means that PWIA results are identical for the two $p$-shells
considered. Moreover, polarization ratios in PWIA may be written in
the general form
\begin{equation}
P'_i = \frac{a_i+b_i\chi+O(\chi^2)}{c_i+d_i\chi +O(\chi^2)},
\kern 1cm 
\mbox{for $i=l,t,n$} \, ,
\end{equation}
where $\chi=p/m_N \sin\theta$, already introduced in~(\ref{rho}-\ref{Jy}),
is the parameter in the SR expansion of
the nuclear current. For low missing momentum, the above fraction has
a linear dependence on $\chi$ plus a small correction of order
$\chi^2$  which breaks linearity for higher $p$. 

For low missing momentum values $p\leq 200$ MeV/c, the effects
introduced by FSI are small, being almost negligible at the maximum of
the momentum distribution ($p\approx 100$ MeV/c).  This result is
expected because of the global reduction of the polarized response
functions produced by FSI: of the order of $\sim30\%$ (fig.~4).  This
is somewhat similar to the behaviour shown by the unpolarized
responses~\cite{Ama99}.  Hence, although not exact because of the
slightly different sensitivities to FSI shown by the various
responses, a kind of cancellation of FSI between the numerator and
denominator in the polarization ratios occurs for low $p$.  From
results in fig.~13, one also observes that FSI effects are slightly
bigger in the case of the $p_{3/2}$ shell, particularly for $P'_l$ and
$\phi=180^{\rm o}$.  The reason of this is connected to the much less
reduction that FSI cause upon the unpolarized $TL$ response for
$p_{3/2}$ (see~\cite{Ama99} for details).

For high missing momentum the DWIA polarizations deviate significantly
from the PWIA results, showing a very pronounced oscillatory behavior
which may even give rise to a change of sign in the polarizations.
This is a clear indication that for high momentum the effects of FSI
are not simply a global reduction of the responses due to the
imaginary part of the potential, but on the contrary, each response
turns out to present a peculiar sensitivity to the interaction. As
shown in figs.~3 and 4, this is hardly visible in the separate
response functions because of the smallness of the momentum
distribution for high $p$. It is important to point out that the
oscillatory behaviour presented by the polarization ratios is a direct
consequence of the breaking of factorization property. This issue was
already studied at the level of the plane wave approach taking care of
the dynamical relativistic effects introduced by the lower components
of the bound Dirac spinors~\cite{Mar02a}.  A general analysis of
factorization within the context of the RDWIA and different
non-relativistic approximations is presently in
progress~\cite{Cris03}.

Focusing on the results presented in fig.~13, we observe that the
shape and magnitude of both polarization asymmetries, $P'_l$ and
$P'_t$, are similar for the two $p$-shells. In the particular case of
$\phi=90^{\rm o}$ (out-of-plane kinematics) the ratio $P'_t$ is very
small, almost negligible for low missing momentum. This is expected
since only the response $W^{T'}_t$, which is very small, contributes
to $P'_t$ in that situation.  For co-planar kinematics a large
discrepancy between the results obtained at $\phi=0^o$ and
$\phi=180^{\rm o}$ exists. As shown by
eqs.~(\ref{RT'}--\ref{Wtgeneral}), the numerator in the ratios $P'_i$,
$i=l,t$, is given through the linear combination
$v_{T'}W^{T'}_i+v_{TL'}W^{TL'}_i\cos\phi$, with the kinematical
factors being $v_{T'}=0.27$ and $v_{TL'}=-0.18$ for the kinematics
considered here (I). Hence, from the transferred polarization
asymmetries measured at $\phi=0^o$ and $\phi=180^{\rm o}$, the
separate responses $W^{T'}_i$ and $W^{TL'}_i$ could be extracted.

Comparing the solid and dashed lines in fig.~\ref{fig-pol0} we
conclude that the uncertainties introduced by the optical potentials
selected are rather small.  For low momentum transfer these
differences are negligible in contrast to fig.~4 where some responses
are shown to be affected appreciably by the optical potential. This
again is an outcome of the fact that the differences between the
responses computed with these potentials are of the same size in
numerator and denominator and they tend to cancel when taking the
quotient to compute the polarizations. Both sets of results start to
differ for $p\geq 300$ MeV/c. Note however that for high $p$-values
other relativistic effects coming from the dynamical enhancement of
the lower components in the wave functions, not included in the
present model, may also contribute significantly to the oscillatory
behaviour of the polarizations~\cite{Mar02a,Cris03}.

The case of the normal polarization transfer $P'_n$
(fig.~\ref{fig-pol00}), present some peculiarities not observed for
$P'_{l,t}$.  First the difference between PWIA and DWIA results is
rather constant for the two shells in the whole range of missing
momentum. In particular, the distorted wave approach leads to results
which are very similar to the ones obtained within PWIA in the case of
the $p_{3/2}$. In addition, the strong oscillatory behaviour due to
FSI and shown for $P'_{l,t}$ (fig.~13) does not appear here, being the
differences introduced by both optical potentials small. 
These results could promote this observable
$P'_n$, that can be obtained in out-of-plane experiments, as a good
candidate in order to study properties of the reaction without being
much affected by FSI.

\subsubsection{Effects of MEC}

\begin{figure}[hp]
\begin{center}
\leavevmode
\def\epsfsize#1#2{0.8#1}
\epsfbox[100 170 500 775]{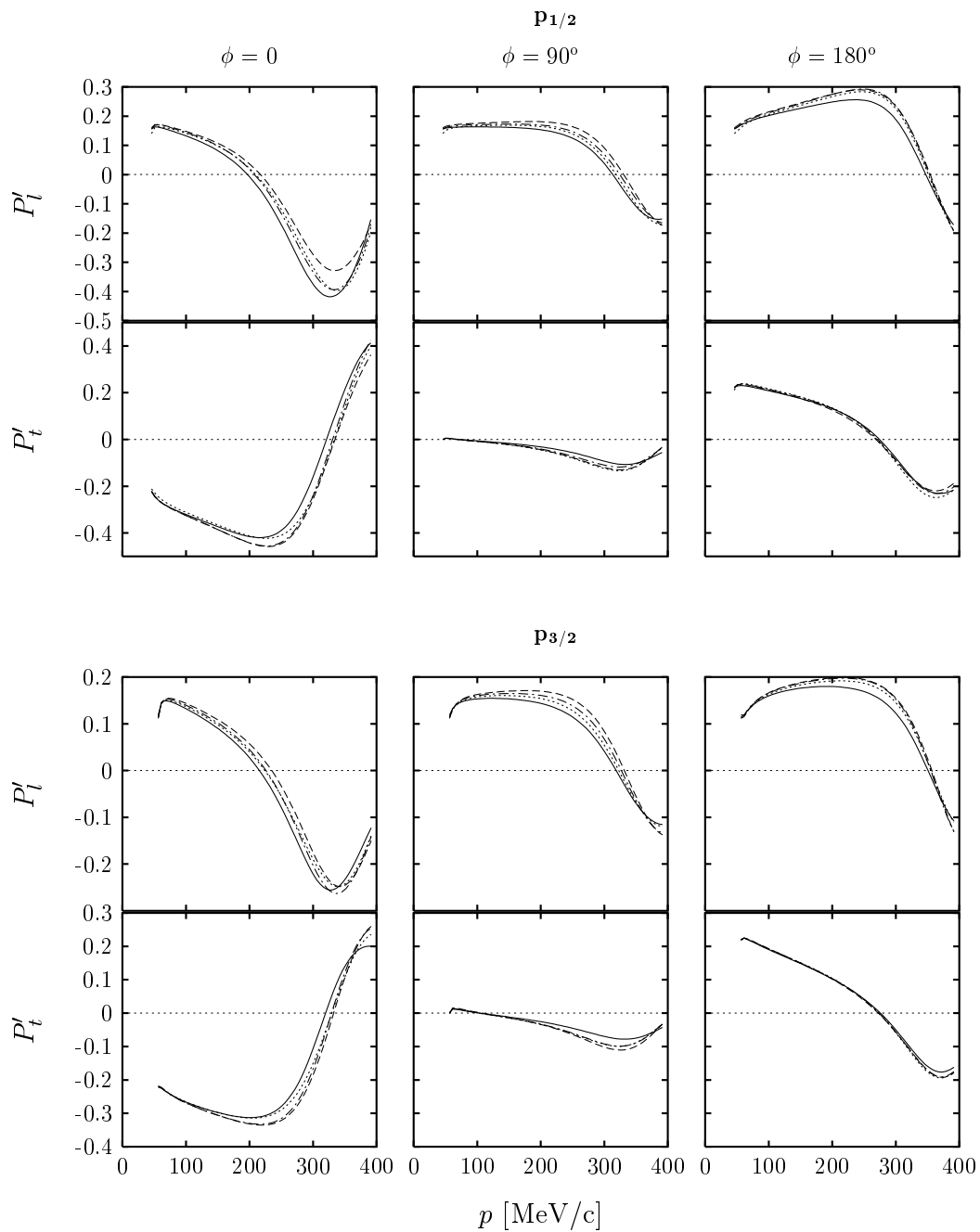}
\end{center}
\caption{ 
\label{fig-pol1}
MEC effects over the transferred polarization asymmetries $P'_l$ and $P'_t$ 
for proton knock-out from the $p$ shells in
$^{16}$O, and $q=460$ MeV/c, $\omega=100$ MeV. The electron
scattering angle is $\theta_e=30^{\rm o}$, and results are shown for
three values of the proton azimuthal angle $\phi=0,90^{\rm o},180^{\rm
o}$. The meaning of the lines is the same as in Fig.~5.}
\end{figure}

\begin{figure}[hp]
\begin{center}
\leavevmode
\def\epsfsize#1#2{0.8#1}
\epsfbox[100 170 500 775]{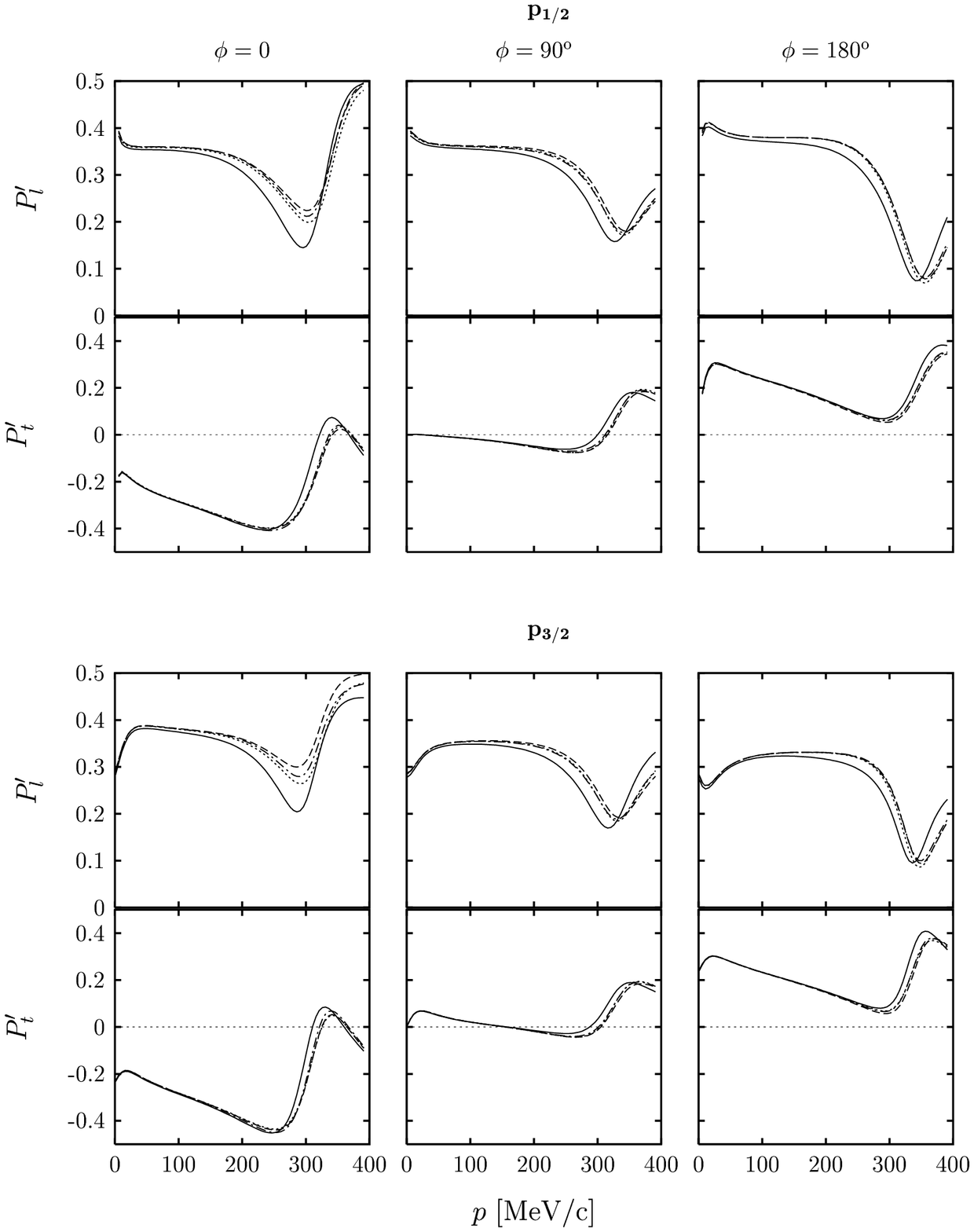}
\end{center}
\caption{
\label{fig-pol2}
The same as Fig.~\ref{fig-pol1} for $q=1000$ MeV/c and $\omega=450$ MeV.
}
\end{figure}

\begin{figure}[hp]
\begin{center}
\leavevmode
\def\epsfsize#1#2{1.0#1}
\epsfbox[100 440 500 785]{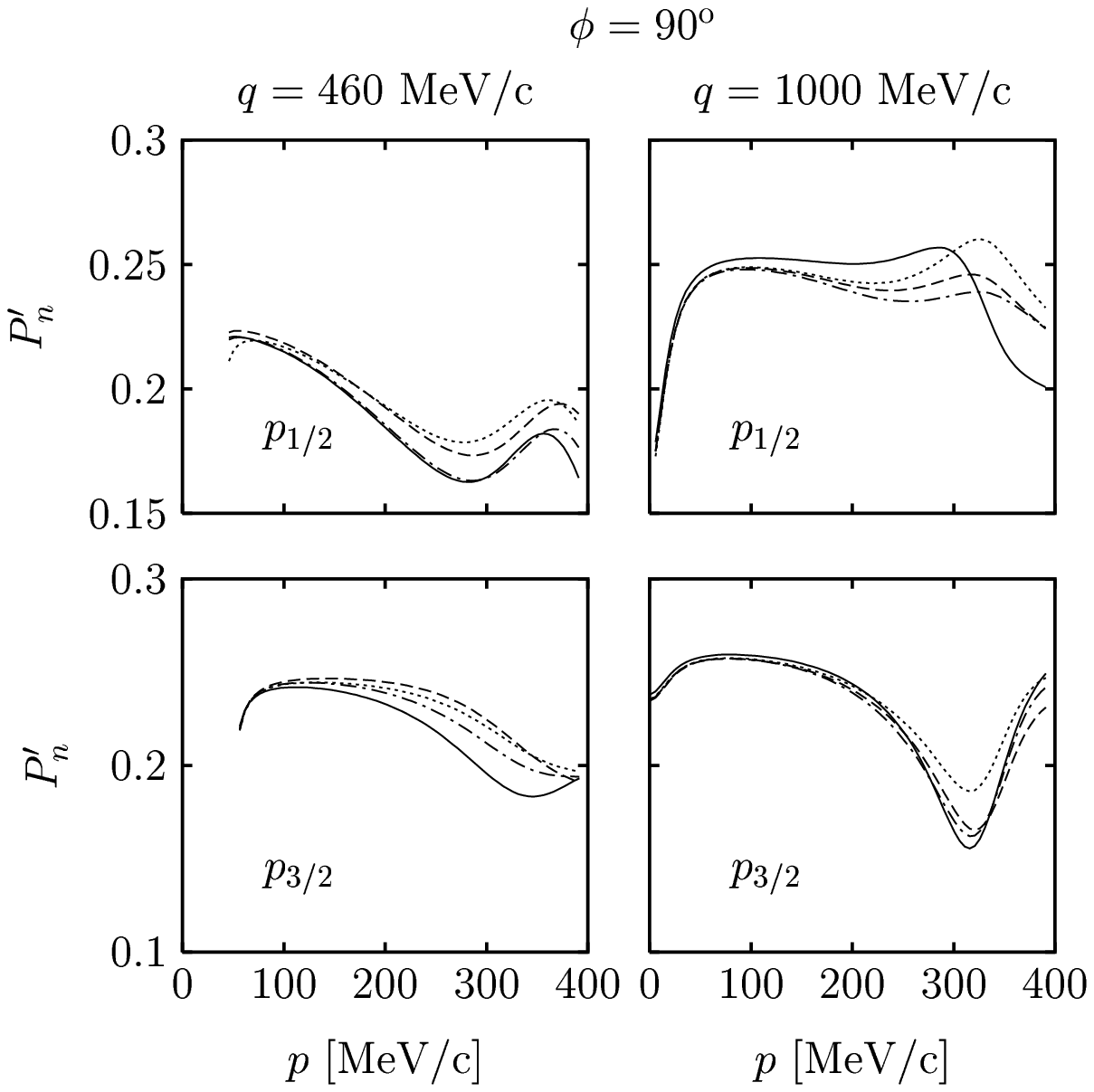}
\end{center}
\caption{
\label{fig-pol3}
The same as Figs.~\ref{fig-pol1} and~\ref{fig-pol2}
for the normal polarization
transfer component $P'_n$ and $\phi=90^{\rm o}$. 
}
\end{figure}

In Figs.~\ref{fig-pol1}--\ref{fig-pol3} we present the effects
introduced by MEC upon the transferred proton polarization components
for the two $p$ shells in $^{16}$O, and for the two kinematics
considered above. In the case of kinematics (I), i.e., $q=460$ MeV/c
(fig.~\ref{fig-pol1}), where we use the optical potential
parameterized by Comfort \& Karp, MEC effects are shown to be small in
the whole missing momentum range and similar for both shells. Note
also that the role played by MEC is of the same order of magnitude or
even smaller than the uncertainty introduced by the use of the two
optical potentials (fig.~\ref{fig-pol0}).  The smallness of MEC
effects on the polarization asymmetries comes from their effective
cancellation when taking ratios of response functions.

Results for higher momentum transfer, $q=1000$ MeV/c (kinematics II),
are shown in Fig.~\ref{fig-pol2} for the Schr\"odinger-equivalent form
of the S-V Dirac global optical potential of ref.~\cite{Coo93}. As in
the previous case, MEC effects are small for low missing momentum;
however they tend to increase significantly for higher $p$-values due
to the $\Delta$ exchange current, inducing a softening of the
transferred polarization asymmetry, which makes its oscillatory
behaviour to appear at slightly lower momenta (see for instance the
important MEC effects observed for $P'_l$ at $\phi=0$, particularly in
the region close to the minimum $p\sim 300$ MeV/c). The present
results indicate that the response functions entering into the
polarization ratios are importantly affected by MEC, mainly due to the
$\Delta$ current, for high momentum transfer.  Large effects of this
kind have also been found recently for the $A_{TL}$ asymmetry obtained
from the analysis of unpolarized (e,e'p) reactions corresponding to
the same kinematics II~\cite{Ama03b}.

The normal polarization $P'_n$ is shown in Fig.~\ref{fig-pol3} for
the two kinematics and both shells. MEC effects
follow the same trends as those observed for $P'_l$ and $P'_t$: they
increase significantly for high momentum transfer ($q=1000$ MeV/c) and high
missing momentum. In contrast with the $P'_l$ and $P'_t$ cases, here
the relative contributions of the separate MEC currents
depend on the specific kinematics and $p$-shell selected, playing the
seagull and pion-in-flight currents an important role.

\begin{figure}[hp]
\begin{center}
\leavevmode
\def\epsfsize#1#2{0.9#1}
\epsfbox[100 280 500 780]{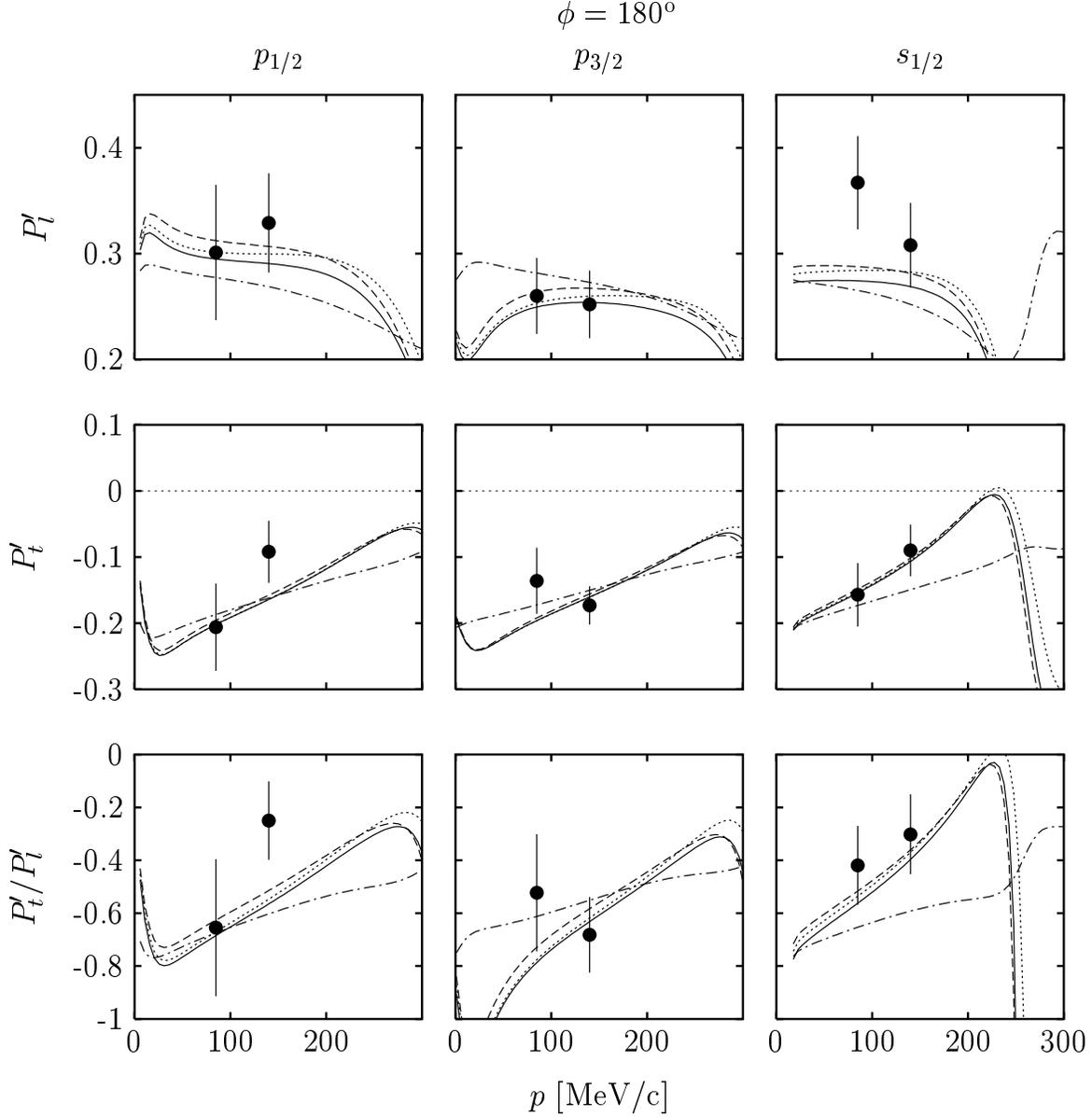}
\end{center}
\caption{ 
\label{fig-pol4}
Transferred polarization asymmetries $P'_l$ and $P'_t$, and
their quotient $P'_t/P'_l$ for proton knock-out from the three shells
in $^{16}$O for $q=1000$ MeV/c and $\omega=450$ MeV. The incident electron
energy is $\epsilon_e=2450$ MeV and the proton azimuthal angle
$\phi=180^{\rm o}$.  The meaning of the lines is the following: dotted
are the DWIA calculation with OB current only; solid are the total
OB+MEC result; dot-dashed are also the OB+MEC result but without the
spin-orbit term of the optical potential. These three curves have been
obtained using the electromagnetic nucleon form factors of Galster. 
Finally, dashed lines are the total OB+MEC result using instead the 
form factors of Gari-Krumplemann. 
}
\end{figure}

To finish we present in Fig.~\ref{fig-pol4} the asymmetries $P'_l$ and
$P'_t$ evaluated for $q=1000$ MeV/c, $\omega=450$ MeV and electron
incident energy $\epsilon_e=2450$ MeV. This kinematics corresponds to
a recent experiment performed at TJlab. We compare our calculations
with the experimental data presented in~\cite{Mal00}. The azimuthal
angle in this experiment was $\phi=180^{\rm o}$. Note the change of
sign of $P'_t$ with respect to the results of Fig.~\ref{fig-pol2}, due
to the opposite definitions of the normal plane (and hence of the $t$
component) for $\phi=180^{\rm o}$ (in our case the normal plane for
$\phi=180^{\rm o}$ would point down in Fig.~1, while in
ref.~\cite{Mal00} it was chosen along the up direction). Results for
the $p$ and $s$ shells in $^{16}$O are shown from left to right.
Although being aware of the possible modifications that the
``dynamical'' relativistic ingredients~\cite{Udi99,Kel99,Mal00} may
introduce in the present calculations, we are rather confident that
the results in fig.~\ref{fig-pol4} give us a clear indication of how
much the DWIA calculation is expected to be modified after including
the two-body (MEC) contributions (compare dotted with solid lines).
As noted, whereas the contribution of MEC over $P'_t$ is negligible,
they give rise to a slight reduction of $P'_l$, which is well inside
the experimental error except for the $s_{1/2}$ shell. Comparing the
results for the two $p$-shells we observe that our model describes
better the case of the $p_{3/2}$. This is in agreement with the
findings in~\cite{Udi99,Ama03b} concerning the $A_{TL}$ asymmetry.
The particular case of $s_{1/2}$ shows that the experimental data for
$P'_t$ are well reproduced by the calculations, which however
understimate very significantly the data for $P'_l$.

In order to clarify the importance of FSI, in fig.~\ref{fig-pol4} we
also show with dot-dashed lines the results corresponding to the
global OB+MEC calculation but without including the spin-orbit term of
the optical potential, i.e., using a phenomenological optical
potential consisting only of a central part. As shown, the
corresponding polarizations present some kind of ``linearity'', being
similar to the PWIA results. This is expected since the spin-orbit
interaction is the main responsible of the oscillatory behaviour of
the polarization ratios, apart from the ``dynamical'' relativistic
effects.

The ratio $P'_t/P'_l$, shown in the bottom panels of
fig.~\ref{fig-pol4}, has been proposed as a suitable observable for
getting information on nucleon properties inside the nuclear
medium~\cite{Arn81}. From inspection of fig.~\ref{fig-pol4} we find
that MEC produce a tiny reduction of this observable, particularly for
low missing momentum, being larger as $p$ goes up.

All of the above results have been computed using the Galster
parameterization for the nucleon form factor. It is of interest to
know the dependence of our results with the nucleon structure, hence
we have also calculated the OB+MEC polarization asymmetries
assuming the Gari-Krumplemann (GK) form factor
parameterization~\cite{Gar85}. The results are shown with dashed
lines. The GK parameterization was used by Udias et al.~\cite{Udi03}
within the context of the relativistic calculations presented in
ref.~\cite{Mal00}. The $P'_l$ computed with GK form factors is
increased with respect to the solid lines, being a little bit closer
to the experimental data.  Let us remind that the effects of MEC lead
to a global reduction of all of these polarization observables, hence
the OB calculation using the GK form factors would be clearly located
above the corresponding results including MEC (dashed lines). This
makes our present results to come closer to the relativistic ones
of~\cite{Mal00}.  Note also that the uncertainty introduced by the
nucleon form factor parameterization shows up in $P'_l$, being
negligible for $P'_t$.

To finish the discussion, it is also interesting to point out that the
behaviour shown by the $P'_t$ data, growing with $p$ for $p_{1/2}$ and
the reverse for $p_{3/2}$, does not agree with the theoretical results
which increase with $p$ for both shells. This is in accordance with
other relativistic calculations~\cite{Mal00}.  
For $p=140$ MeV/c our predictions for $P'_t$ in the case of the 
$p_{1/2}$ shell clearly underestimate the data; as already mentioned,
other relativistic effects coming from the lower components of the Dirac
wave functions, not considered here, may also play a significant role.


\subsection{Comparison with previous works}

Concerning previous calculations of MEC in $(e,e'p)$ reactions, in~\cite{Ama99,Ama03b} 
comparisons for unpolarized observables obtained
with the present model with those of Refs.~\cite{Bof91,Ryc01,Slu94,Giu02} were presented.  
Next we summarize the differences of MEC effects on recoil polarization
observables between the present work and Refs.~\cite{Bof90,Ryc99}.

\begin{enumerate}

\item  Boffi and collaborators~\cite{Bof90} find for intermediate $q$ values large MEC effects 
on the separate polarized responses (reduction of the order of 15-30\% or
even larger), being the $\Delta$ current the main contribution. We
get in general smaller and qualitatively different 
effects for this kinematics, the seagull and
pion in flight currents being in our case as important as the $\Delta$.
Concerning the transfer polarization ratios, they find
$P'_l$ to be the most sensitive one, with a 20\% decrease due to the 
$\Delta$ for low missing momentum $p<200$ MeV/c. 
In our calculation MEC effects are clearly less important for these missing momentum values.


\item Ryckebusch {\em et al.}~\cite{Ryc99} do not present the separate $l,t,n$ 
response functions. In general they find small MEC effects, as we do, in the transferred
polarizations for low $p<300$ MeV/c. These effects being larger as
$q$ and $p$ increase. Comparing specifically our
results to theirs for kinematics II, we observe that the OB results clearly differ
due to the different treatment of FSI, while somewhat larger and
qualitatively different MEC effects are found in this work.

\end{enumerate}

Be it as it may, since the different treatment of FSI 
and of the current operators in~\cite{Bof90,Ryc99} and in the present work
produces discrepancies already at the impulse approximation, it is hard
to draw general conclusions on MEC effects beyond the fact
that in~\cite{Bof90} MEC lead to excesively large contributions compared with us, while
their small size in~\cite{Ryc99} is in accord with our calculation.


\section{Summary and conclusions}


In this paper we have presented a distorted wave model of
$(\vec{e},e'\vec{p})$ reactions which goes beyond the impulse
approximation with the inclusion of two-body meson exchange currents.
Relativistic kinematics to relate the energy and momentum of the
ejected proton is used and the currents are derived through an
expansion in powers of the missing momentum over the nucleon
mass. Explicit expressions of the polarized response functions in a
general multipole expansion method are given. Results for the
responses and transferred polarization asymmetries have been obtained
for proton knock-out from the different shells in $^{16}$O for
quasiperpendicular kinematics with the transfer momentum fixed to
$q=460$ and 1000 MeV/c.  FSI have been considered in each case by
using different optical potentials.

One of our primary goals has been to estimate, within our present
model, the validity of the impulse approximation by analyzing the
effect of MEC on the different recoil nucleon polarized
observables. Thus we compare the standard DWIA results, obtained using
only the OB current, with the ``full'' calculation which includes the
MEC.  We have also explored the role played by the particular
description of the FSI, hence we compare the results obtained by using
different optical potentials which have been widely considered in the
literature: Schwandt~\cite{Sch82} and Comfort \& Karp~\cite{Com80}
parameterizations. For higher energy we have used instead the
Schr\"odinger-equivalent form of a Dirac optical potential.

From our present studies we may summarize and conclude the following:

\begin{enumerate}

\item The induced $T$, $TL$ and $TT$ polarized responses are
      particularly sensitive to the details of the optical potential,
      allowing them, specially the $TT$ ones, to constrain the
      theoretical model for FSI. The transferred polarized responses
      ($T'$ and $TL'$), which survive in PWIA, show a much less
      sensitivity to the interaction.

\item In general, MEC effects over the transferred $T'$, $TL'$ 
      polarized responses
      for $q=460$ MeV/c and moderate missing momentum ($p<300$ MeV/c)
      are rather small ($< 5\%$), and tend to increase as $q$ goes
      higher, being of the order of a $\sim 10\%$ reduction (due
      mainly to the $\Delta$ current) in the particular case of
      $W^{T'}_l$ and $q=1$ GeV/c.

      The role of MEC gets clearly more important for the induced $T$,
      $TL$ and $TT$ polarized responses. Emphasis should be placed on
      $W^{TT}_t$ and $W^{TT}_n$ which are reduced at the maximum by
      $\sim 20\%$ and $\sim 30\%$, respectively, for $q=460$ MeV/c and
      for the $p_{1/2}$ shell; notice however that these effects are
      negligible in the case of $p_{3/2}$. For $q=1$ GeV/c the role of
      MEC diminishes.

\item FSI give rise to an important deviation of the transferred
      polarization asymmetries $P'_l$ and $P'_t$ with respect to the
      PWIA results, showing a very pronounced oscillatory behaviour
      that starts for $p\geq 200$ MeV/c. This behaviour does not
      appear in the component $P'_n$. The uncertainties introduced by
      the optical potentials are rather small for the missing momentum
      region analyzed.

\item MEC effects on $P'_t$ and $P'_l$ are negligible for $q=460$
      MeV/c and increase for $q=1$ GeV/c, especially for $p>200$
      MeV/c.  The role of MEC on $P'_n$ is clearly more important.

\end{enumerate}

Finally we are confident that the significant sensitivity shown by
some polarized observables to MEC, particularly to the $\Delta$
current, will be maintained within the scheme of a ``fully''
relativistic calculation which takes care of relativistic ingredients,
such as the dynamical enhancement of lower components, not included in
the present model. Work along this line is in progress.

\section*{Acknowledgments}
This work was partially supported by funds provided by DGI (Spain) and
FEDER funds, under Contracts Nos BFM2002-03218, BFM2002-03315 and
FPA2002-04181-C04-04 and
by the Junta de Andaluc\'{\i}a.


\appendix 
\section{Sum over third components and reduced response functions}


In this appendix we perform the sum over third components in the
multipole expansion of response functions and give their explicit
expressions in terms of the reduced matrix elements of the multipole
operators.

First we write the response functions~(\ref{rlrt}--\ref{rt'rtl'}) 
in terms of the spherical components of the current matrix elements 
$J_{\pm1}\equiv \langle\np' \ns, B|\hat{J}_{\pm1}|A\rangle$ using
the hadronic tensor~(\ref{hadronic-tensor})
\begin{eqnarray}
R^L &=& \frac{1}{K} 
\sum \rho^*\rho \\
R^T &=& \frac{1}{K} \sum \left(|J_{-1}|^2+|J_{+1}|^2\right)\\
R^{TL} &=& -2\frac{1}{K}  {\rm Re} \sum \rho^*\left(J_{+1}-J_{-1}\right)\\
R^{TT} &=& \frac{1}{K} \sum \left(J_{-1}^*J_{+1}+J_{+1}^*J_{-1}\right)\\
R^{TL'} &=& -2 \frac{1}{K} {\rm Re} \sum \rho^*\left(J_{+1}+J_{-1}\right)\\
R^{T'} &=& \frac{1}{K} \sum \left(|J_{+1}|^2-|J_{-1}|^2\right)\, .\\
\end{eqnarray}
Inserting the multipole expansion for the charge and current
components as given in~(\ref{m-rho},\ref{m-J}) we find that each 
response can be written as a sum of terms of the type
\begin{equation}
B^{m'm}_{J'J}
\equiv \frac{1}{K}
\sum_{M_B}
\langle \np'\ns,J_BM_B| \hat{T'}_{J'm'}(\nq)|A\rangle^*
\langle \np'\ns,J_BM_B| \hat{T}_{Jm}(\nq)|A\rangle \, ,
\end{equation}
where $\hat{T}'_{J'm'}$ and $\hat{T}_{Jm}$ represent in general the
Coulomb, electric or magnetic multipole operators.  Introducing now
the multipole expansion~(\ref{final-states}) corresponding to the
final state, which is polarized along an arbitrary direction
$\ns$~(\ref{rotation}), we get
\begin{eqnarray}
B^{m'm}_{J'J}
&=& \frac{1}{K}
\sum_{M_B}\sum_{\nu'\nu}
{\cal D}^{(1/2)}_{\nu'\frac12}(\ns)
{\cal D}^{(1/2)}_{\nu\frac12}(\ns)^*
\nonumber\\
&&
\kern -1cm \times
\sum_{l'M'j'm'_p}
i^{l'} Y^*_{l'M'}(\hp')
\langle \frac12\nu' l' M'| j'm'_p\rangle
\langle j' m'_p J_BM_B|J'm'\rangle
\langle (l'j')J_B,J'm'|T'_{J'm'}|A\rangle^*
\nonumber\\
&&
\kern -1cm \times
\sum_{lMjm_p}
i^{-l} Y_{lM}(\hp')
\langle \frac12\nu l M| jm_p\rangle
\langle j m_p J_BM_B|Jm\rangle
\langle (lj)J_B,Jm|T_{Jm}|A\rangle \, ,
\end{eqnarray}
where we have used $J_f=J$ and $M_f=m$, since the initial nucleus has 
total angular momentum equal zero. 

Using the Wigner-Eckart theorem for the matrix elements of tensor operators 
between states of definite angular momenta
\begin{equation}
\langle (lj)J_B,Jm|T_{Jm}|0\rangle
=
\frac{1}{[J]}\langle (lj)J_B,J\|T_{J}\|0\rangle
\end{equation}
and reducing the products of two rotation matrices and two spherical
harmonics to linear combinations of
spherical harmonics
\begin{eqnarray}
{\cal D}^{(1/2)}_{\nu'\frac12}(\hs)
{\cal D}^{(1/2)}_{\nu\frac12}(\hs)^*
&=&
\sqrt{4\pi}\sum_{\Jb\Mb}(-1)^{1/2+\nu+\Jb}f^{(1/2)}_{\Jb}
\tresj{\frac12}{\frac12}{\Jb}{-\nu}{\nu'}{\Mb}
Y_{\Jb\Mb}(\hs)
\\
Y^*_{l'M'}(\hp')Y_{lM}(\hp')
&=&
\nonumber\\
&&
\kern -1.5cm
\sum_{\Jb'\Mb'}(-1)^M\frac{[l][l'][\Jb']}{\sqrt{4\pi}}
\tresj{l}{l'}{\Jb'}{-M}{M'}{\Mb'}
\tresj{l}{l'}{\Jb'}{0}{0}{0}
Y_{\Jb'\Mb'}(\hp'),
\nonumber\\
\end{eqnarray}
with $f^{(1/2)}_{\Jb}=\frac{1}{\sqrt{2}}$ the Fano tensor for 
spin-1/2 polarization, we obtain
\begin{eqnarray}
B^{m'm}_{J'J}
&=&
\sum_{M_B}\sum_{\nu\nu'}\sum_{\Jb\Mb}
\sum_{lMjm_p}\sum_{l'M'j'm'_p}\sum_{\Jb'\Mb'}
i^{l'-l}(-1)^{\frac12+\nu+\Jb}f^{(\frac12)}_{\Jb}
\tresj{\frac12}{\frac12}{\Jb}{-\nu}{\nu'}{\Mb}
Y_{\Jb\Mb}(\ns)
\nonumber\\
&&
\mbox{}\times
(-1)^M[l][l'][\Jb']
\tresj{l}{l'}{\Jb'}{-M}{M'}{\Mb'}
\tresj{l}{l'}{\Jb'}{0}{0}{0}
Y_{\Jb'\Mb'}(\hp')
\nonumber\\
&&
\mbox{}\times
(-1)^{l'-\frac12-m'_p}[j']\tresj{\frac12}{l'}{j'}{\nu'}{M'}{-m'_p}
(-1)^{J_B-j'-m'}\tresj{j'}{J_B}{J'}{m'_p}{M_B}{-m'}
\nonumber\\
&&
\mbox{}\times
(-1)^{l-\frac12-m_p}[j]\tresj{\frac12}{l}{j}{\nu}{M}{-m_p}
(-1)^{J_B-j-m}\tresj{j}{J_B}{J}{m_p}{M_B}{-m}
\nonumber\\
&&
\mbox{}\times
\langle (l'j')J_B,J'\|T'_{J'}\|0\rangle^*
\langle (lj)J_B,J\|T_{J}\|0\rangle \, ,
\label{B}
\end{eqnarray}
where we have transformed the Clebsch-Jordan to three-J coefficients.
Next we perform the sums over third components of angular momenta in
the above expression. Note that the total phase inside the sum can
be simplified to
\begin{equation}
\mbox{phase}= (-1)^{\frac12+m_p}(-1)^{\Jb+l+l'}(-1)^{j-j'}.
\end{equation}
Therefore the following coefficient appears
\begin{eqnarray}
S &\equiv& 
\sum_{M_B}\sum_{\nu\nu'}\sum_{MM'}\sum_{m_pm'_p}
(-1)^{\frac12+m_p+\Jb+l+l'+j-j'}
\tresj{\frac12}{\frac12}{\Jb}{-\nu}{\nu'}{\Mb}
\tresj{l}{l'}{\Jb'}{-M}{M'}{\Mb'}
\nonumber\\
&&
\kern -1.5cm
\mbox{}\times
\tresj{\frac12}{l'}{j'}{\nu'}{M'}{-m'_p}
\tresj{j'}{J_B}{J'}{m'_p}{M_B}{-m'}
\tresj{\frac12}{l}{j}{\nu}{M}{-m_p}
\tresj{j}{J_B}{J}{m_p}{M_B}{-m} \, .
\nonumber\\
\label{S}
\end{eqnarray}
We first perform the sum over $\nu,\nu',M,M'$
by using a 9-j coefficient
\begin{eqnarray}
\lefteqn{
\sum_{\nu\nu'}\sum_{MM'}
\tresj{\frac12}{\frac12}{\Jb}{-\nu}{\nu'}{\Mb}
\tresj{l}{l'}{\Jb'}{-M}{M'}{\Mb'}
\tresj{\frac12}{l}{j}{\nu}{M}{-m_p}
\tresj{\frac12}{l'}{j'}{\nu'}{M'}{-m'_p} =}
\nonumber\\
&&
(-1)^{\frac12+l+j}\sum_{LM}[L]^2
\tresj{\Jb}{\Jb'}{L}{\Mb}{\Mb'}{M}
\tresj{L}{j}{j'}{M}{m_p}{-m'_p}
\nuevej{\Jb}{\Jb'}{L}{\frac12}{l}{j}{\frac12}{l'}{j'} \, .
\end{eqnarray}
Next we compute the sum over $m_p,m'_p,M_B$ using a 6-j coefficient
\begin{eqnarray}
\lefteqn{\sum_{m_pm'_pM_B}
(-1)^{m_p+\frac12}
\tresj{L}{j}{j'}{M}{m_p}{-m'_p}
\tresj{j}{J_B}{J}{m_p}{M_B}{-m}
\tresj{j'}{J_B}{J'}{m'_p}{M_B}{-m'} =}
\nonumber\\
&&
(-1)^{\frac12+m'+j+j'+J_B}
\tresj{L}{J'}{J}{M}{-m'}{m}
\seisj{L}{J'}{J}{J_B}{j}{j'}\, .
\end{eqnarray}
Then the $S$-coefficient (\ref{S}) results
\begin{eqnarray}
S &=&
\sum_{LM}[L]^2(-1)^{\Jb+l'+j+J_B+m'}
\tresj{L}{J'}{J}{M}{-m'}{m}
\tresj{\Jb}{\Jb'}{L}{\Mb}{\Mb'}{M}
\nonumber\\
&&
\mbox{}\times
\nuevej{\Jb}{\Jb'}{L}{\frac12}{l}{j}{\frac12}{l'}{j'}
\seisj{L}{J'}{J}{J_B}{j}{j'}\, .
\label{S-final}
\end{eqnarray}
To finish we insert the result~(\ref{S-final}) into~(\ref{B}), and
define indices $\sigma,\sigma'$ corresponding to the quantum numbers
of the final states
\begin{equation} 
\sigma = (l,j,J), \kern 1cm
\sigma' = (l',j',J') 
\end{equation}
and a coupling coefficient
\begin{eqnarray}
\Phi_{\sigma'\sigma}(\Jb,\Jb',L)
&=&
\sqrt{2}[l][l'][j][j'][J][J'][\Jb'][L]
(-1)^{l+j+J_B+L+J+J'}
\nonumber\\
&&
\mbox{}\times
\tresj{l}{l'}{\Jb'}{0}{0}{0}
\seisj{L}{J'}{J}{J_B}{j}{j'}
\nuevej{\Jb}{\Jb'}{L}{\frac12}{l}{j}{\frac12}{l'}{j'}\, .
\label{phi}
\end{eqnarray}
The final expression for $B$ is
\begin{eqnarray}
B^{m'm}_{J'J}
&=&
\frac12\sum_{lj}\sum_{l'j'}\sum_{\Jb\Jb'LM}
i^{l'-l}
(-1)^m
\tresj{J}{J'}{L}{m}{-m'}{M}
\Phi_{\sigma'\sigma}(\Jb,\Jb',L)
\nonumber\\
&&
\mbox{}\times
\left[ Y_{\Jb}(\hs) Y_{\Jb'}(\hp')\right]_{L,-M}
\langle (l'j')J_B,J'\|T'_{J'}\|0\rangle^*
\langle (lj)J_B,J\|T_{J}\|0\rangle \, ,
\label{Bfinal}
\end{eqnarray}
where the coupling between two spherical harmonics has been used
\begin{eqnarray}
\lefteqn{\left[Y_{\Jb}(\hs)Y_{\Jb'}(\hp')\right]_{LM}=}
\nonumber\\
&=& (-1)^{\Jb-\Jb'+M}\sum_{\Mb\Mb'}[L]
\tresj{\Jb}{\Jb'}{L}{\Mb}{\Mb'}{-M}
Y_{\Jb\Mb}(\hs)Y_{\Jb'\Mb'}(\hp')\, .
\label{Ycoupling}
\end{eqnarray}

Although the result given in eq.~(\ref{Bfinal}) is formally identical,
with the exception of the factor $1/2$, to the one obtained
in~\cite{Ama98b} for the case of polarized nuclei, there exists a
basic difference concerning the polarization coefficient
$\Phi_{\sigma'\sigma}(\Jb,\Jb',L)$, which contains all the information
on the polarization properties of the particles in the initial and/or
final state. Note that in the present case (spin-1/2 polarized
particles), the angular momentum in the expansion of the rotation
matrices, $\Jb$, only takes the values 0,1.  The case $\Jb=0$ is the
only one surviving when the final nucleon is not polarized, i.e., when
summing the cross sections for $\pm s$ values.  In this case the
present formalism reduces simply to the standard unpolarized one of
ref.~\cite{Maz02}. In fact, for $\Jb=0$ we have $\Jb'=L$ and the
reader can prove after some Racah algebra, that
$\Phi_{\sigma'\sigma}(0,L,L)$ reduces to the expression given in
eq.~(A11) of Ref.~\cite{Maz02} for the unpolarized case.

Moreover, using the properties of the 9-j symbol, the following 
important symmetry property is found for the polarization coefficient
under exchange of the indices
\begin{equation}
\Phi_{\sigma'\sigma}(\Jb,\Jb',L)= (-1)^{\Jb+\Jb'+L}
\Phi_{\sigma\sigma'}(\Jb,\Jb',L).
\end{equation}
This property coincides with the one already presented
in~\cite{Ama98b} in the case of polarized targets. Since the multipole
expansion of response functions performed in~\cite{Ama98b} was based
on this symmetry, then the same formalism can be applied to the
present case. In this way one arrives to
eqs.~(\ref{WL11}--\ref{WTL'1M}) (see~\cite{Ama98b} for more details on
the expansion).


\end{document}